\documentclass[12pt]{article}
\usepackage{eqlist}
\usepackage{amssymb}
\usepackage{dsfont}
\usepackage{mdwlist}
\usepackage{amsmath}
\usepackage{graphicx}
\usepackage{subfigure}
\usepackage{eqlist}
\usepackage{mathrsfs}
\normalsize
\begin{document}
\addtolength{\baselineskip}{.20mm}
\newlength{\extraspace}
\setlength{\extraspace}{2mm}
\newlength{\extraspaces}
\setlength{\extraspaces}{2mm}

\newcommand{\newsection}[1]{
\vspace{15mm} \pagebreak[3] \addtocounter{section}{1}
\setcounter{subsection}{0} \setcounter{footnote}{0}
\noindent {\Large\bf \thesection. #1} \nopagebreak
\medskip
\nopagebreak}
\newcommand{\newsubsection}[1]{
\vspace{1cm} \pagebreak[3] \addtocounter{subsection}{1}
\addcontentsline{toc}{subsection}{\protect
\numberline{\arabic{section}.\arabic{subsection}}{#1}}
\noindent{\large\bf 
\thesubsection. #1} \nopagebreak \vspace{3mm} \nopagebreak}
\newcommand{\ba}{\begin{eqnarray}
\addtolength{\abovedisplayskip}{\extraspaces}
\addtolength{\belowdisplayskip}{\extraspaces}

\addtolength{\belowdisplayshortskip}{\extraspace}}

\newcommand{\be}{\begin{equation}
\addtolength{\abovedisplayskip}{\extraspaces}
\addtolength{\belowdisplayskip}{\extraspaces}
\addtolength{\abovedisplayshortskip}{\extraspace}
\addtolength{\belowdisplayshortskip}{\extraspace}}
\newcommand{\ee}{\end{equation}}
\newcommand{\STr}{{\rm STr}}
\newcommand{\figuur}[3]{
\begin{figure}[t]\begin{center}
\leavevmode\hbox{\epsfxsize=#2 \epsffile{#1.eps}}\\[3mm]
\parbox{15.5cm}{\small
\it #3}
\end{center}
\end{figure}}
\newcommand{\im}{{\rm Im}}
\newcommand{\calm}{{\cal M}}
\newcommand{\call}{{\cal L}}
\newcommand{\sect}[1]{\section{#1}}
\newcommand\hi{{\rm i}}
\def\bea{\begin{eqnarray}}
\def\eea{\end{eqnarray}}

\makeatletter
\newcommand{\rmnum}[1]{\romannumeral #1}
\newcommand{\Rmnum}[1]{\expandafter\@slowromancap\romannumeral #1@}
\makeatother

\begin{titlepage}
\begin{center}

\vspace{3.5cm}

{\Large \bf{Reconstruct the Distance Duality  Relation by  Gaussian Process}}\\[1.5cm]

{Yi Zhang $^{a,}$\footnote{Email: zhangyia@cqupt.edu.cn}} \vspace*{0.5cm}

{\it $^{a}$College of Mathematics and Physics, Chongqing University
of Posts and Telecommunications, \\ Chongqing 400065, China
}

\date{\today}
\vspace{3.5cm}

\textbf{Abstract} \vspace{5mm}

 \end{center}
 In this letter,  the distance-duality  (DD)  relation is reconstructed   by Gaussian process (GP) which  is cosmological  model-independent.  Generally, the GP plays two important roles.  One is  to shape the $\eta$  tendency which denotes the deviation from the DD relation, the other one is   to produce the luminosity-distance (LD, $D_{L}$) and the angular-diameter-distance (ADD, $D_{A}$) data at the same redshift. 
 The shapes of $\eta$ are given out  based on  SNe Ia (Type Ia supernovae) data with different light-curve fitters (including MLCS2K2 and SALT2) and ADD data with different galaxy cluster morphologies (including the elliptical $\beta$ and spherical $\beta$ models).   The   data related to  MLCS2K2 light-curve fitter  have higher values of $\eta$ compared to  that related to the SALT2 light-curve fitter. 
As for the morphology of galaxy cluster, the DD relation is favored by the elliptical one.

\end{titlepage}

\section{Introduction}\label{sec1}
 In astronomy, the distance-duality (DD) relation  is of fundamental importance as it connects  the luminosity-distance (LD, $D_{L}$) and the angular-diameter-distance (ADD, $D_{A}$).   
  Initially, the DD relation ($D_{L}/D_{A}(1+z)^{2}=1$)  is  called   Etherington relation \cite{Tolman,Etherington}.  It  
   is valid in any cosmological  background where photons travel along null geodesics and where 
  photon number is conserved. In fact,  many geometric properties are invariant when the roles of source and observer in astronomical observations are transposed. As the key element of    analyzing    the  galaxies observations, the Cosmic Black-body Radiation (CBR) observations  and  the gravitational lensing,
  DD relation  has been usually used in cosmology   for granted. 
    A major crisis would arise  for observational cosmology if  DD relation  was  found to not be true.
    However,  if  there were deviations from a metric theory of gravity or variations of fundamental constants, or  photons not traveling along unique null geodesics,  the DD relation 
  would be violated  theoretically \cite{Ellis1}.

In the earlier work \cite{Uzan:2004my,Bassett:2003vu,
Lima:2011ye,Avgoustidis:2010ju,DeBernardis:2006ii,Holanda:2012at,Goncalves:2013kxh,Holanda:2010ay,Nair:2012dc,Jhingan:2014hfa,Nair:2011dp,Lazkoz:2007cc,Lampeitl:2009jq},  a new parameter  
\begin{eqnarray}
\eta=\frac{D_{L}}{D_{A}(1+z)^{2}},
\end{eqnarray} 
    is defined to    denote the departure from  DD relation.   $\eta$ was tested as constant and parameterizations \cite{Cao:2011pb,Cao:2011fw, Holanda:2011hh,Li:2011exa,Meng:2011nt,Yang:2013xpa,Holanda:2010vb}. 
                Maximum likelihood estimation is employed to determine the most probable values which needs $\eta$ sample first.   One   way of getting $\eta$ sample  is to combine the  distance results from both observation and theoretical expressions.    Obviously, the test depends on  cosmological model heavily. 
    The LD or  ADD data could    be fixed by the $\Lambda$CDM model with  the fixed observational cosmological parameters. 
Observationally, the LD data could be derived from Type Ia supernovae (SNe Ia) compilation; while the ADD data could be estimated from astrophysical sources such as Baryon acoustic oscillation (BAO) data,  galaxy clusters, FRIIb galaxies,radio galaxies, \emph{etc.}.  However, the main problem for the observational-only data is that the observational LD and ADD data are not at the same redshift. Therefore, the observational data must be combined with the theoretical one.
Anyway,  an interesting contradiction appeared: Refs.\cite{DeBernardis:2006ii,Lazkoz:2007cc,Lampeitl:2009jq,Nair:2011dp} saw no evidence of violation for DD relation  
while a violation of the DD relation is concluded by Ref.\cite{Bassett:2003vu}.

In recent work,  local regression methods are used (\emph{e.g.} Ref.\cite{Nair:2012dc}) to avoid  introducing errors due to the redshift mismatch. Specially, 
the method of  binning LD data  with the redshift range $\delta z=|z_{LD}-z_{ADD}|<0.005$    worth attentions. In Refs.\cite{Li:2011exa,Meng:2011nt,Holanda:2010vb},   the DD relation is found to  discriminate the morphology of cluster galaxy. For example,  Li \emph{et~al.} \cite{Li:2011exa}  found that the DD relation can be accommodated at $1\sigma$ confidence level (CL) for the elliptical one, and at $3\sigma$ CL for the spherical  one.

   Though the data sample based on the $\delta z=|z_{LD}-z_{ADD}|<0.005$ criterion  is model-independent, the derived  $\eta$ is still model-dependent.  In principle, the DD relation should  be tested only from the astronomical observations, \emph{i.e.}
finding cosmological sources whose intrinsic luminosity and intrinsic sizes are known. 
Gaussian process (GP) is  a powerful non-linear interpolation tool which  allows one to reconstruct a function from data without assuming a parameterization \cite{gp1,gp2,gp3,gp4}.  It  has been applied to cosmology for the equation of state of dark energy \cite{Seikel:2012uu,Nair:2013sna,Holsclaw:2011wi,Holsclaw:2010sk,Holsclaw:2010nb,Suzuki:2011hu,Amanullah:2010vv,Busti:2014aoa}.  
   In this letter, the Gaussian process will be applied  to   reconstruct the DD relation which  will play two main roles: to get  the shape of $\eta$ and to match the redshifts of the LD  and  ADD data. 
  SNe Ia data from different light-curve fitters will be investigated.  Specificlly, we will use  the SNe Ia compilations  in Ref.\cite{Kessler:2009ys,Jha:2006fm}  which  were derived  from both SALT2 and MLCS2K2 light-curve  fitters by Kessler \emph{et~al.}  and the Union2.1 compilation  based on the SALT2 fitter  \cite{Guy:2005me}. 
Furthermore,     the   ADD sample with different morphological models   are chosen as well. The elliptical  \cite{DeFilippis:2005hx,Reese:2002sh,Mason} and the  spherical galaxy clusters \cite{Bonamente:2005ct} are used to test the intrinsic shape of cluster.

The letter is organized as follows. In Sec.\ref{sec2}, the  GP will be introduced. 
 In Sec.\ref{sec3},   we will give out the luminosity-distance and  the angular-diameter-distance.  
 In Sec.\ref{sec4}, the GP is used to get the LD and  ADD data at the same redshift. In
 Sec.\ref{sec5}, we will use new $\eta$ samples to test the DD relation, one is based the improved criterion  that the redshift difference of 
 ADD and  LD should be less than $0.001$.  The other one is based on  GP data of the  LD and the observational ADD data  (or GP data of the ADD and the observational LD data, or  both GP data of the LD and ADD). 
 The  deviation from   $\eta=1$ will be  investigated. And  a summary will be given out in Sec.\ref{sec6}.


\section{Gaussian Process}\label{sec2}
 As a powerful non-linear interpolating tool,  Gaussian process is useful for cosmology and astronomy \cite{gp1,gp2,gp3,gp4}. 
The GP has been introduced as non-parametric technique and has successfully reconstructed  the equation of state  of dark energy \cite{Seikel:2012uu,Nair:2013sna,Holsclaw:2011wi,Holsclaw:2010sk,Holsclaw:2010nb}.
Here,  the public available  code GAPP (Gaussian Processes in Python) \cite{Seikel:2012uu,Nair:2013sna}  is used  to reconstruct the DD relation. 
In the following, if GP was applied,  the label ``(GP)" would be put after the name of the data.

 A Gaussian distribution is a distribution over random variables, while a GP is distribution over functions  which has elements of an infinite dimensional 
  space. 
 Here,  we  give the basic conception of GP  which could perform a reconstruction of  function from data without assuming function parameterization. 
 In the reconstruction of DD relation,  the observational  data and the reconstructed data are combined to  form a joint distribution. Based on the joint distribution,   the GP is given by a mean function with Gaussian error bands, where the function values at different points  are connected through a covariance function. 
 
The freedom in  GP comes from the chosen covariance function which determines how smooth the process is and how nearby points are correlated.
Throughout the whole letter, we will use the  Matern covariance function
  \begin{eqnarray}
 k(z,\tilde{z}) = \sigma_{f}^{2}\frac{2^{1-\mu}}{\Gamma(\mu)}\frac{\sqrt{2\mu(z-\tilde{z})^{2}}}{l}K_{\mu}(\frac{\sqrt{2\mu(z-\tilde{z})^{2}}}{l})
 \end{eqnarray}
where $z$ are $\tilde{z}$ represent two different points,  hyper-parameter    $\sigma_f$  is  the typical change of  variable, hyper-parameter $l$ is the typical length scale of function,  $\Gamma$ is the Gamma function and  $K_{\mu}$ is a modified Bessel function with $\mu=5/2,7/2$ and $9/2$.   The hyper-parameters $\sigma_f$ and $l$ could be trained by maximizing the marginal likelihood which 
 only depends on the locations of the observations. Indeed,  the GP approach is rather robust to the covariance function choice \cite{gp1,gp2,gp3,gp4}.

\section{Observational Data}\label{sec3}
 The deviation from  $\eta=1$  corresponds to the violation of   DD relation.  Even a  small deviation 
 indicates  breakdown of fundamental physical theories. 
 It is obvious that  the observational  ADD  
 and  LD data  are needed  if we want to reconstruct  DD relation. In isotropic and homogenous cosmological background,  the angular-diameter-distance   and luminosity-distance,
 \begin{eqnarray}
 \label{d}
&&D_A= \frac{1}{ (1+z)} \int_{0}^{z}\frac{dz'}{H(z')},\,\,\,\,\  D_L= (1+z)\int_{0}^{z}\frac{dz'}{H(z')},
\end{eqnarray}
 are also called standard rulers and standard distances where $H$ is the Hubble parameter.

\subsection{The Angular-Diameter-Distance Data}
 Two   galaxy cluster data  sets  which provide the ADD data based on different  morphology and dynamics are  used here.

The  first  samples are formed by 25 galaxy clusters in  redshift range of $0.023\leq z \leq 0.784$ \cite{DeFilippis:2005hx,Reese:2002sh,Mason}.
  In Ref. \cite{DeFilippis:2005hx}, cosmological angular-diameter-distance $D_A|_{cosm}$  is known from the redshift and the prior knowledge of the cosmology.    Based on Eq.(10) in Ref.\cite{DeFilippis:2005hx} and assuming  the isothermal triaxial $\beta$ cluster profile, the experimental quantity of ADD  was calculated which combined X-ray surface brightness and SZE analysis.  We call it the  ES$|_{ell}$ sample instead  of the name  $D_c|_{exp}^{ell}$ in Ref.\cite{DeFilippis:2005hx}.
 And, Ref.\cite{DeFilippis:2005hx} also listed  the experimental estimate of  SS$|_{circ}$ which was assumed with spherical symmetry and reported by Refs. \cite{Reese:2002sh,Mason}.
  The $D_A|_{cosm}$ data will be used as  standard rulers. The ES$|_{ell}$ will be used for morphology comparison. The  SS$|_{circ}$ data will be used for data comparisons.

The second samples  are defined by the 38 ADD galaxy clusters from Ref.\cite{Bonamente:2005ct} where the cluster plasma and dark matter distributions were analyzed by  assuming generalized spherical $\beta$ model.  We call  it   the   Spherical  Sample (SS). They are in  the redshift range of $0.14 \leq z \leq 0.89$.
As reported by Ref.\cite{Bonamente:2005ct}, almost all the ADD data  are followed by asymmetric uncertainties. To deal with the asymmetric uncertainties, we choose three data dealing methods for comparison  and call them  SS\Rmnum{1} , SS\Rmnum{2}  and SS\Rmnum{3}. The first one  addressed this issue by combining the statistical and systematic uncertainties in quadrature  \cite{Holanda:2010ay,Li:2011exa}: 
\begin{eqnarray}
&&SS\Rmnum{1} : \,\,\,\ \emph{E}(D_{A}) = D_{A} ,\,\,\,\  \sigma_{D_A}=\sqrt{\sigma_{+}^{2}+\sigma_{-}^{2}}.
\end{eqnarray}
where the $\emph{E}(D_{A})$  denotes  the expected value,  $\sigma_{D_{A}}$ is the standard deviation, $\sigma_{+}$ and $\sigma_{-}$ are the upper and lower limits of error.  
We also use the  reported $D_{A}$ value as expected value  and the larger flank of each two-sided error  as standard deviation   \cite{Meng:2011nt}
 \begin{eqnarray}
 SS\Rmnum{2} : \,\,\,\ \,\,\,\   \emph{E}(D_{A}) = D_{A}, \,\,\,\, \sigma_{D_A}= max(\sigma_{+}, \sigma_{-}). 
\end{eqnarray}
As stressed by Ref.\cite{Bonamente:2005ct}, the modeling uncertaintis of the angular diameter distances presented contribute to statistical uncertainties. Therefore, it is reasonable to make the following correction and estimations \cite{D'Agostini:2004yu} , 
\begin{eqnarray}
\label{ss3}
 SS\Rmnum{3} : \,\,\,\  \emph{E} (D_{A})=D_{A}+0.75(\sigma_{+}-\sigma_{-}) ,\,\,\,\,  \sigma _{D_A} = \frac{\sigma_{+}+\sigma_{-}}{2}. 
\end{eqnarray}
Specially, the error of $D_A|_{cosm}$ and SS$|_{circ}$ data are asymmetrical too. For simplity, they will be dealt by  the third method.

\subsection{The Luminosity-distance Data}
   In order to check consistency of GP, we will perform  same analysis among SNe Ia samples which have different light-curve fitters.  We consider the  MLCS2K2  (the current version of  Multicolor light-curve Shape method) and SALT2 (the current version of Spectral Adaptive light-curve Template)  light curve fitters which are
 the commonly used.
 The two different  light-curve  fitters reflect different assumptions about the nature of color variations in SNe Ia, different approach  to training the models using pre-existing data, and  different ways of determining parameters.

The SNe Ia data provide the luminosity-distance  relation
\begin{equation}
\label{dl}
\mu_{B}(z)  =5 \log_{10}\frac{D_{L}(z)}{1Mpc}+25,
\end{equation}
where $\mu_{B}$ is the distance moduli. 
Anyway,   the SALT2  fitter which is a  spectral-template-based fit method \cite{Guy:2005me}  depends on the cosmological parameter where the cosmological parameters and light-curve parameters are fitted simultaneously. In contrast, there is no such   cosmological parameter dependence for the MLCS2K2 fitter \cite{Kessler:2009ys,Jha:2006fm}.

In Ref.\cite{Kessler:2009ys},  the SN Ia data is  analyzed  with both  MLCS2K2  and SALT2 light-curve fitters. It has  288  data set of distance moduli in the redshift range of $0.024\leq z \leq 1.390$ with a 247  data set in the redshift range of $0.01\leq z \leq0.8$.  They are dubbed as MLCS2K2:09 and SALT2:09. The  Union2.1 SNe Ia compilation \cite{Suzuki:2011hu}  has been analyzed with the SALT2 light-curve fitter and  has 580 dataset  of $\mu_{B}(z)$ in the redshift range of $0.01\leq z \leq 1.41$ with a 508  dataset in the redshift range of $0.01\leq z \leq0.8$.  
In principle, the data related to Union2.1 and that related to SALT2:09 should  give out similar results.  

\subsection{Our GP}

We give out the details of how we use data with GP.  Besides the luminosity-distance and the angular-diameter-distance,  the value of $z$ is  required  for getting  $\eta$. 
 As the above section mentioned, the data set  of  Union2.1 (or MLCS2K2:09) is  larger than that of galaxy cluster.  To be precise,  we use the redshift in the luminosity-distance data to obtain $\eta$.
 Technically, we transfer the SNe Ia distance moduli to a new variable $D_{L}(1+z)^{-2}$.  
 
And,   we choose the range of reconstructed redshift  as  $0\leq z \leq 0.8$ which will cover the our smallest   redshift region $0.023\leq z \leq 0.784$.  
   And  as the observational redshift is in the precision of $0.001$, 
 the redshift number of GP is chosen as 800. Thus, every observational data in $0\leq z \leq 0.8$ will correspond to a reconstructed one.
 
 At last, as Ref.\cite{Seikel:2012uu} suggested, we have run   Gaussian process repeatedly with different initial conditions for the hyper-parameters ($\sigma_f$, $l$) to get a smooth result in case that  the optimization of the hyper-parameters  getting stuck in a local maximum.

\section{Rough Checking of  GP}\label{sec4}


\begin{figure}  \centering
  {\includegraphics[width=2.0in]{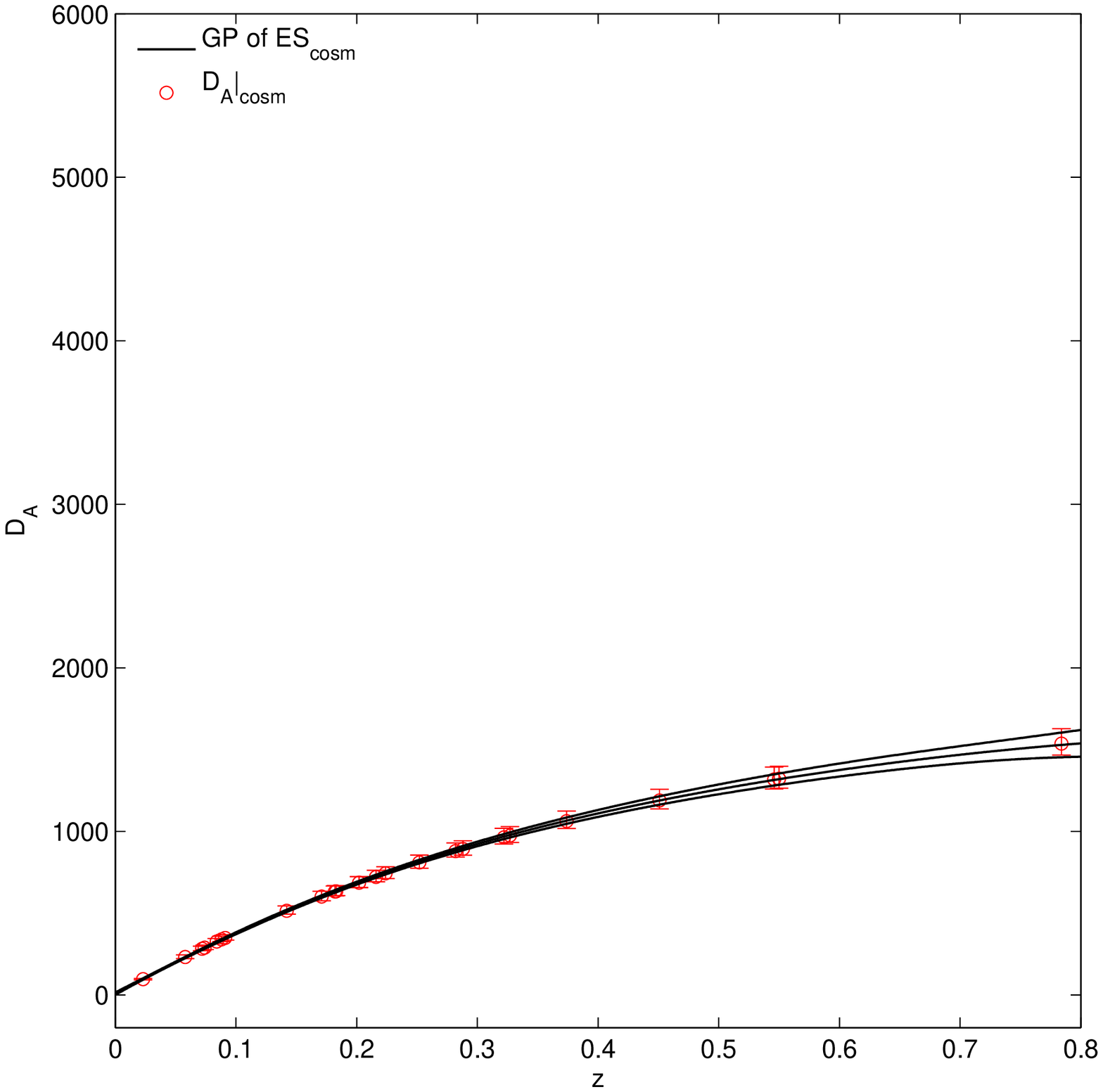}}\quad
    {\includegraphics[width=2.0in]{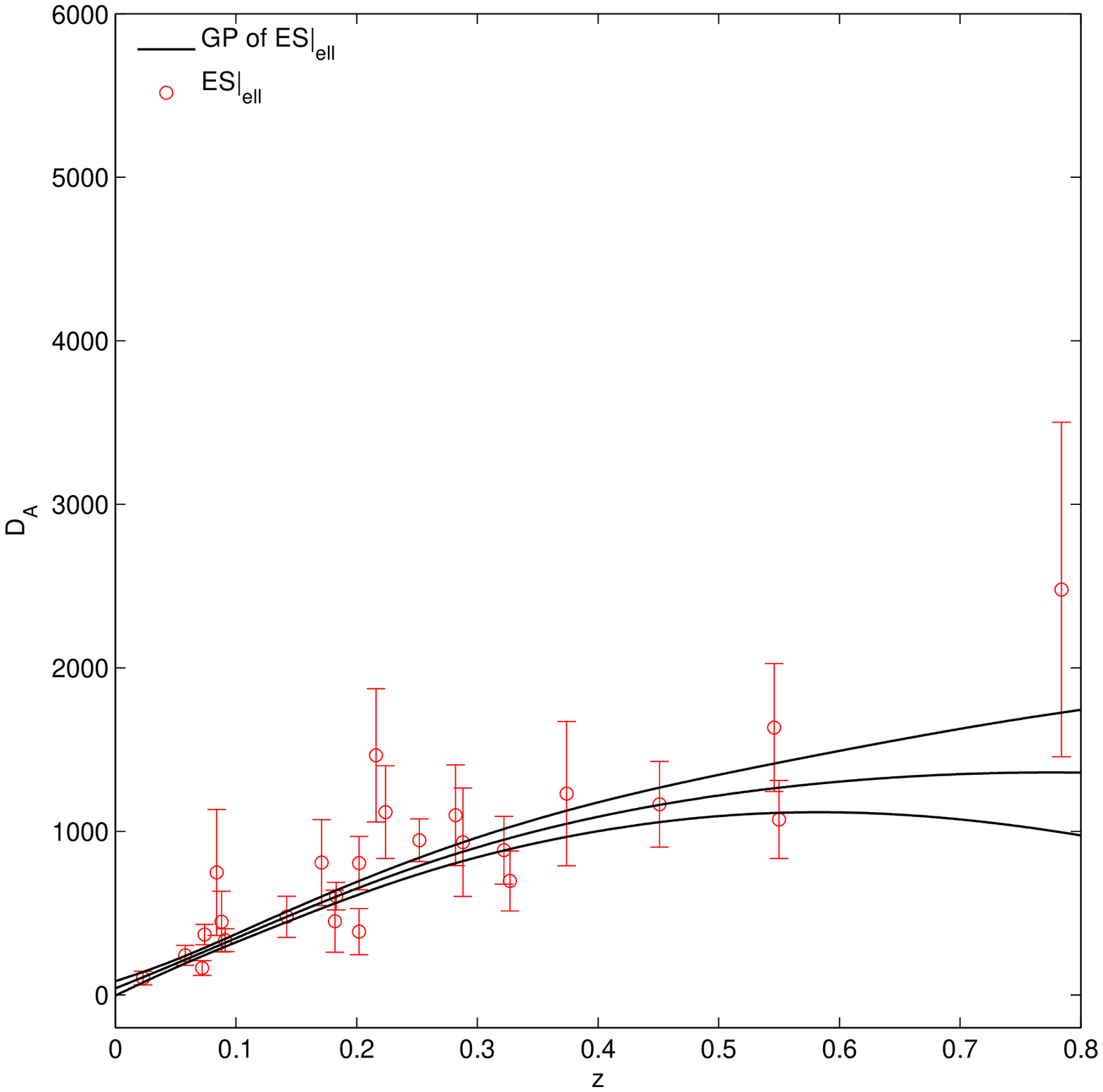}}\quad 
        {\includegraphics[width=2.0in]{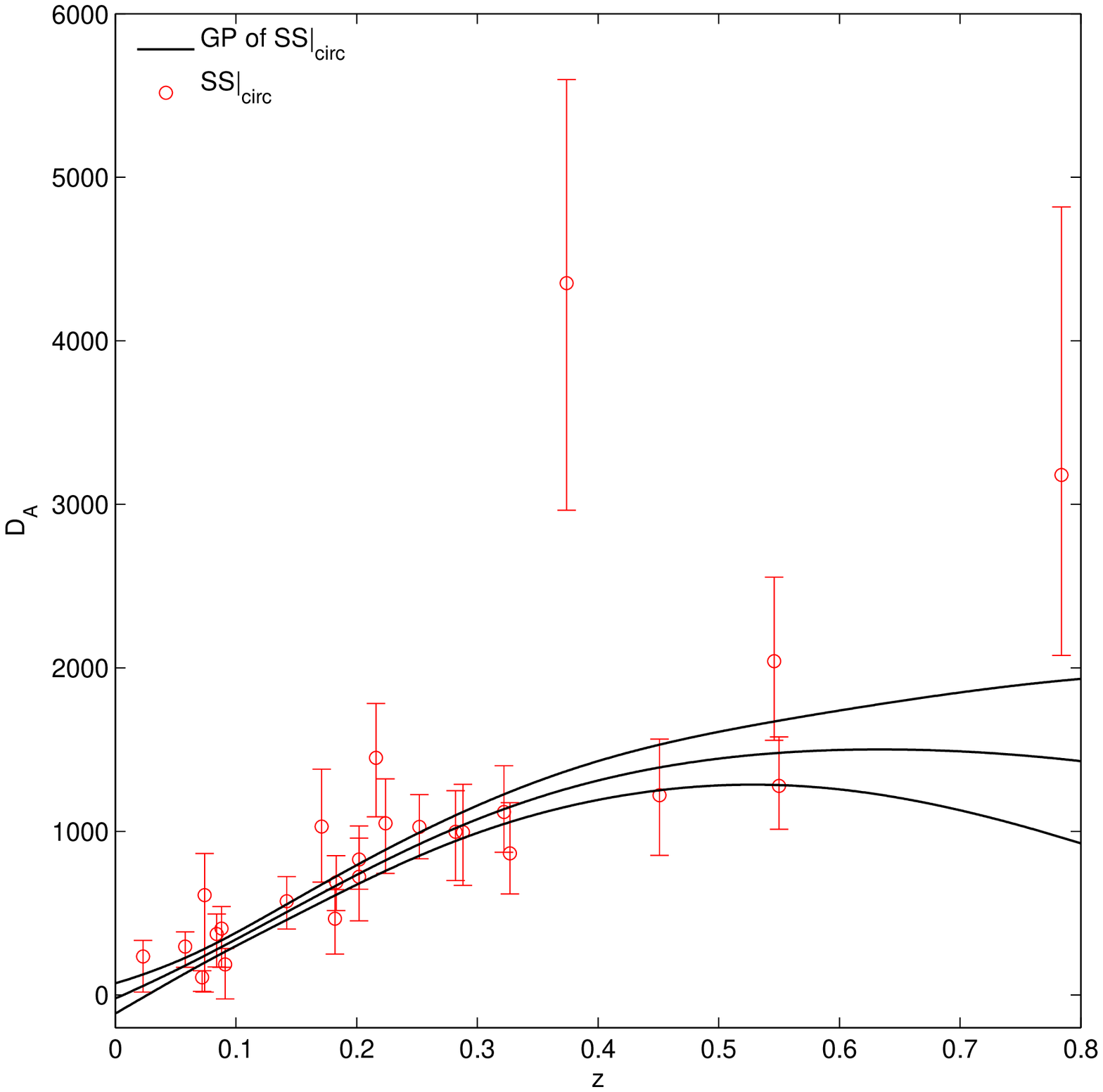}}\quad \\
  {\includegraphics[width=2.0in]{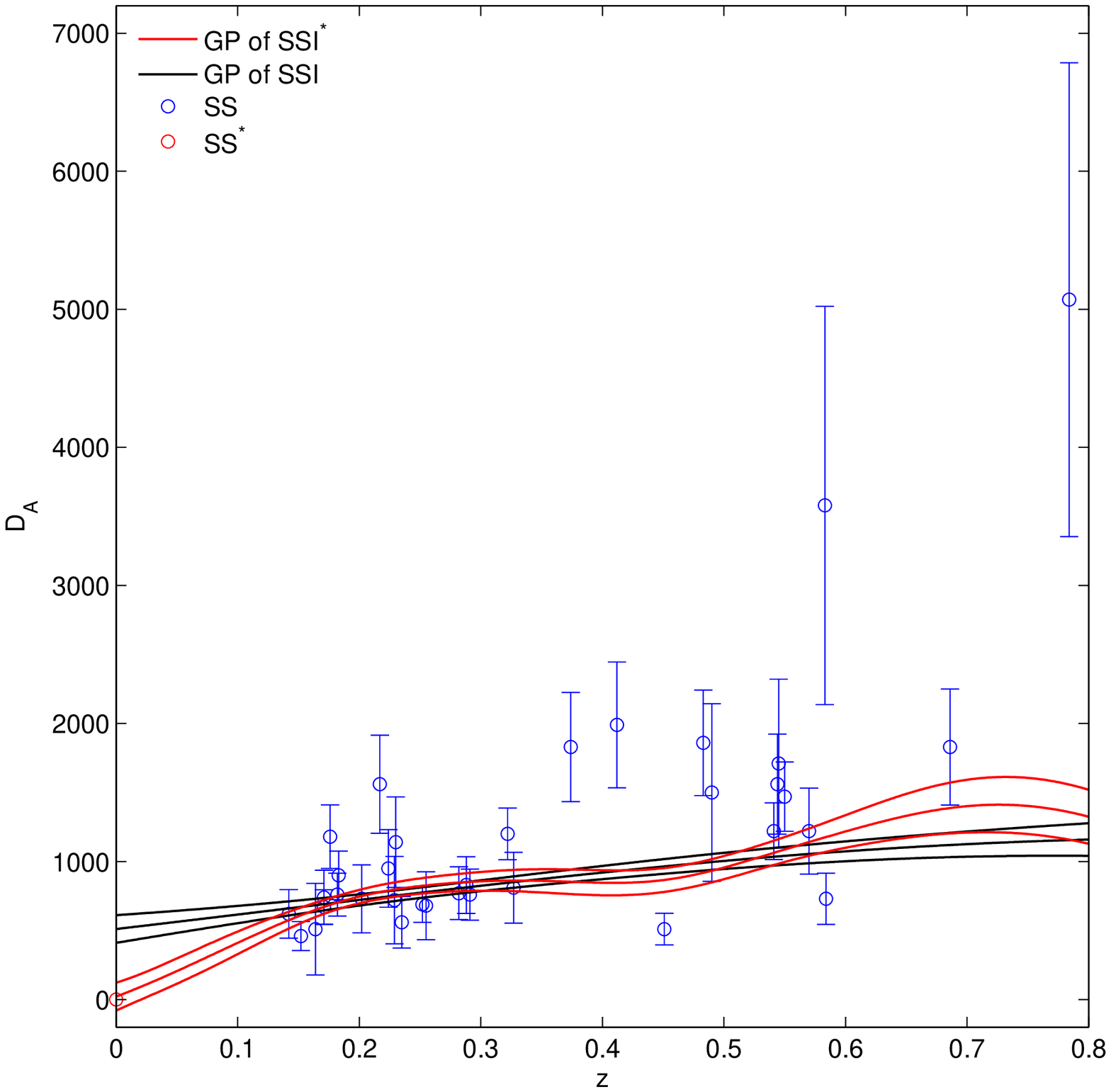}}\quad
    {\includegraphics[width=2.0in]{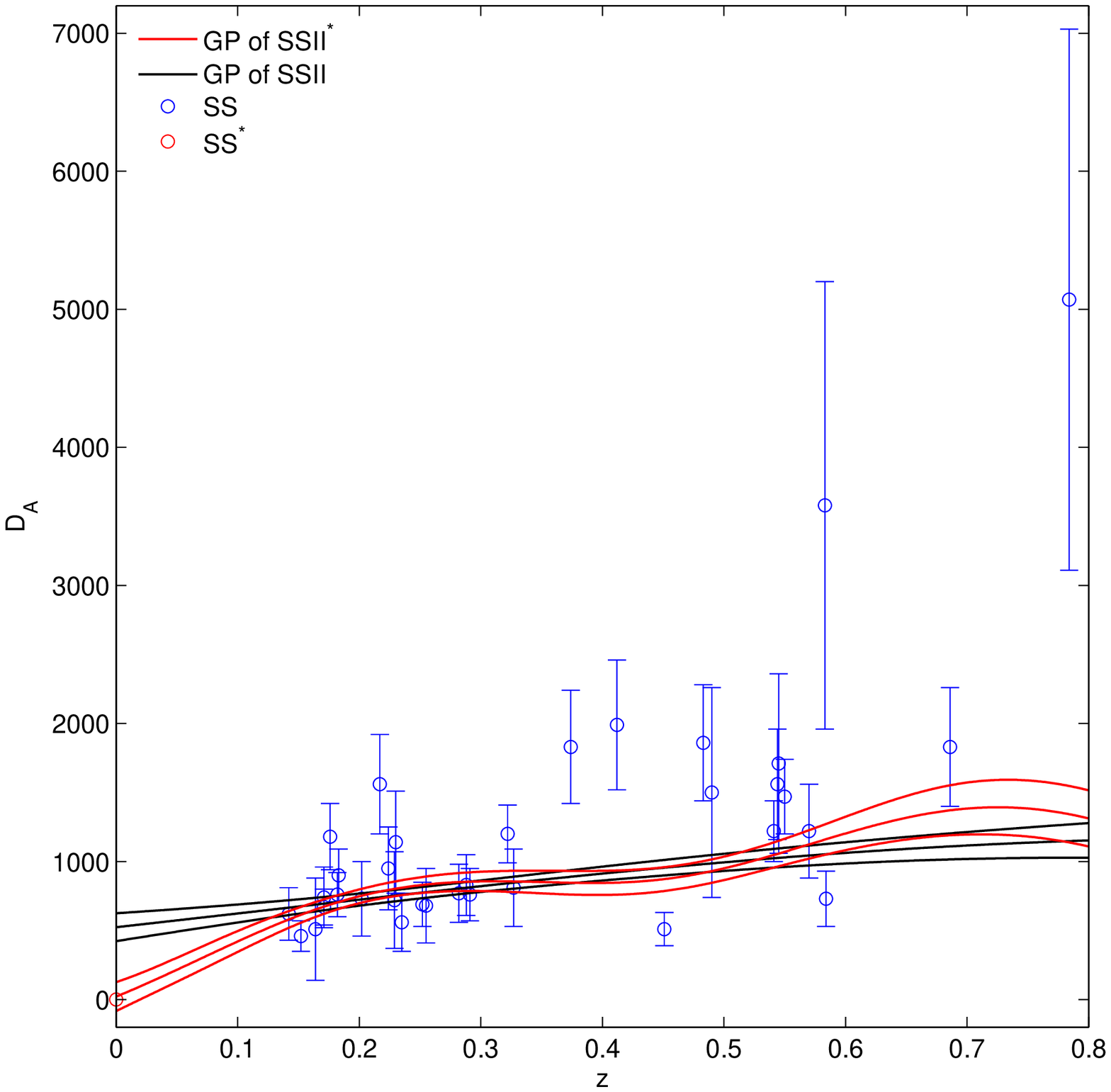}}\quad
                      {\includegraphics[width=2.0in]{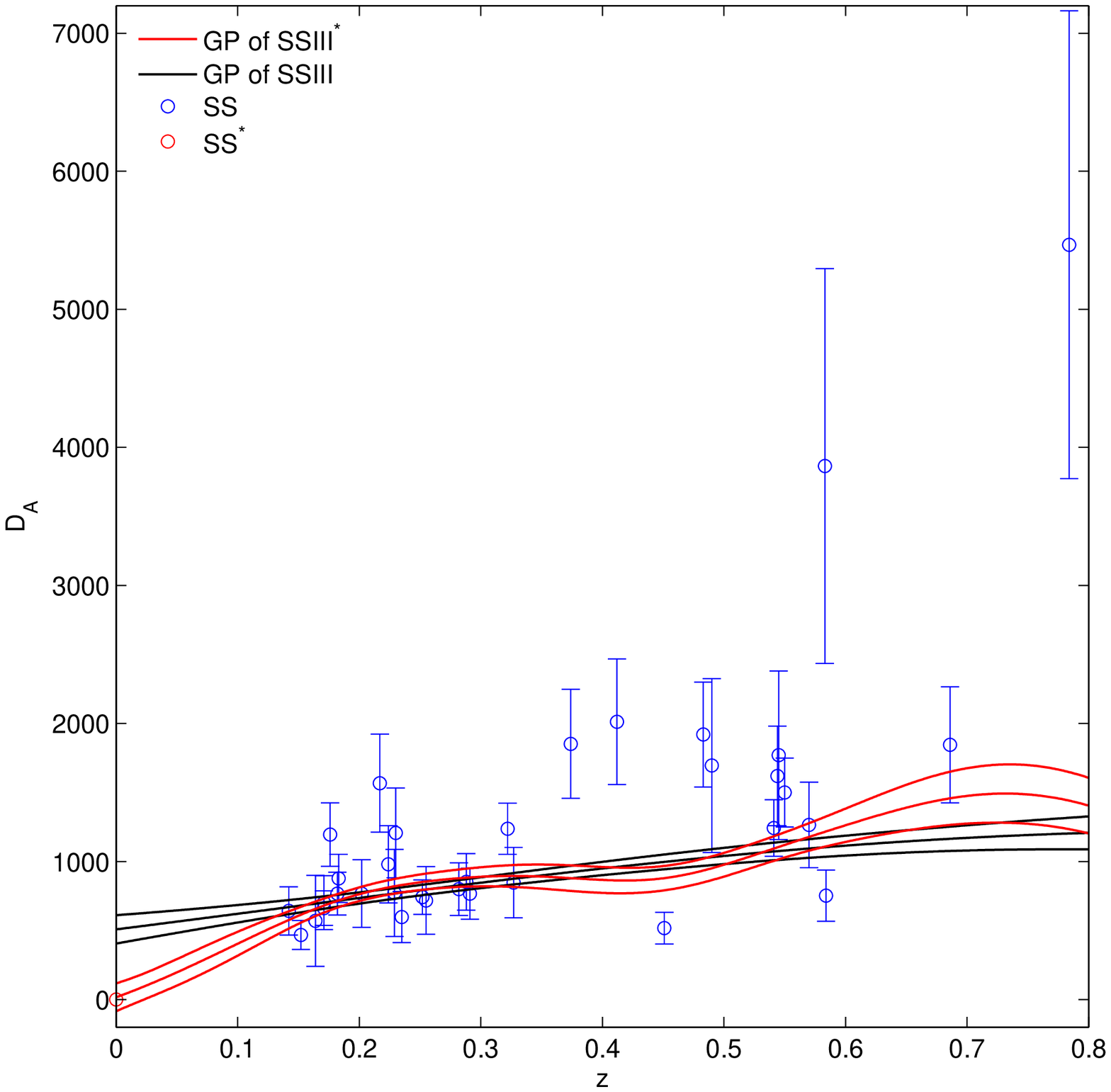}}\quad\\
  {\includegraphics[width=2.0in]{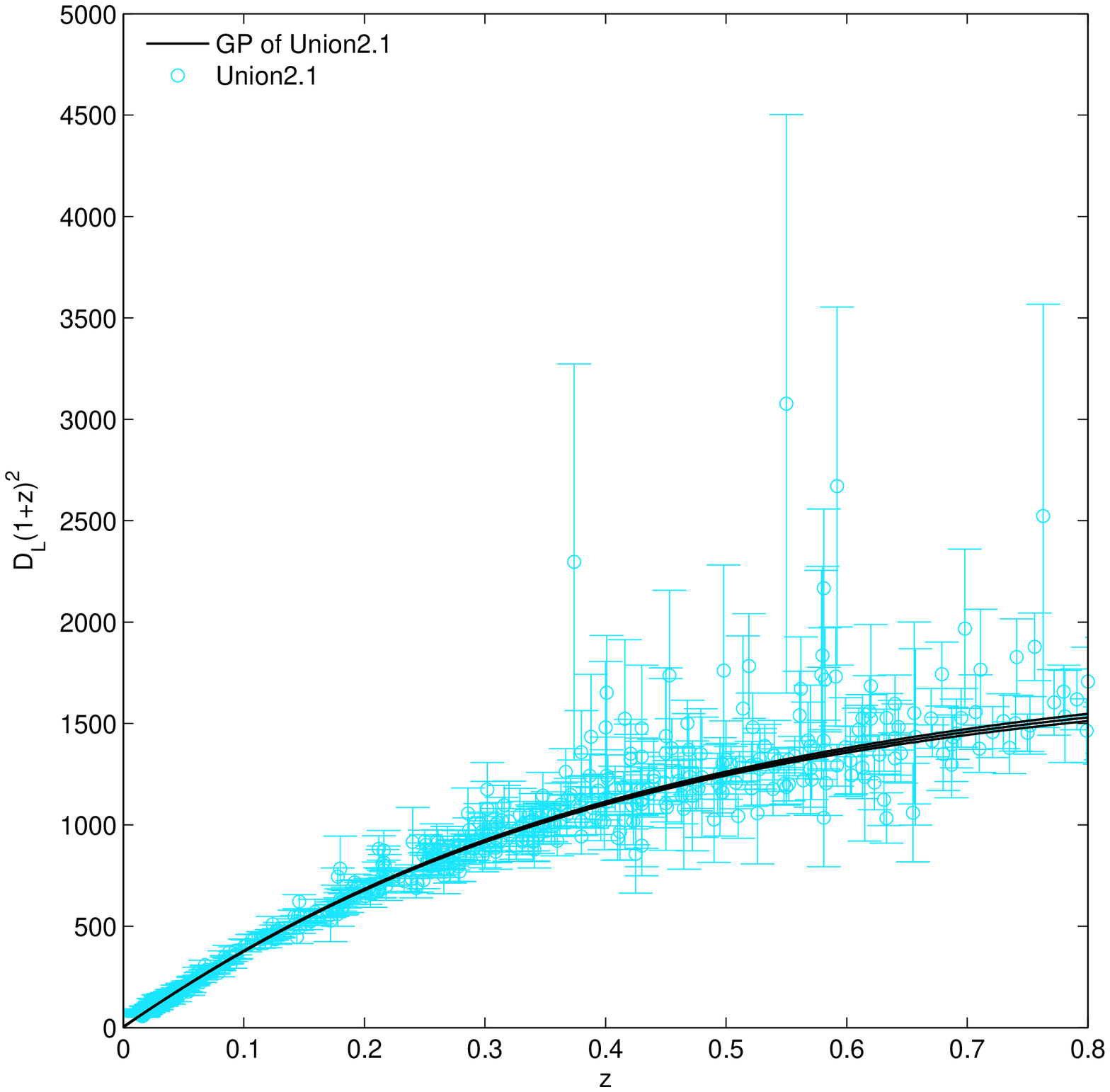}}\quad
    {\includegraphics[width=2.0in]{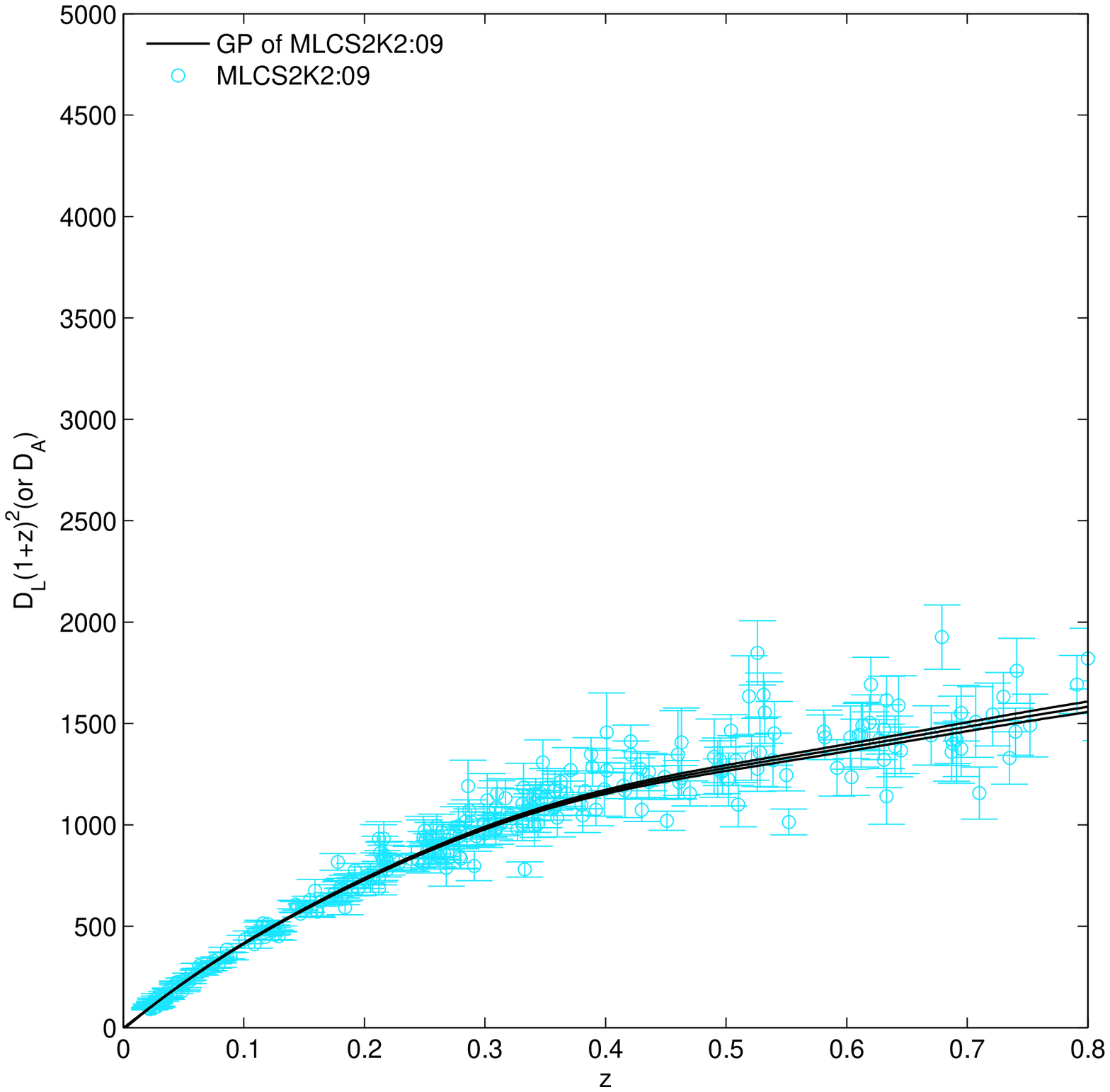}}\quad
     {\includegraphics[width=2.0in]{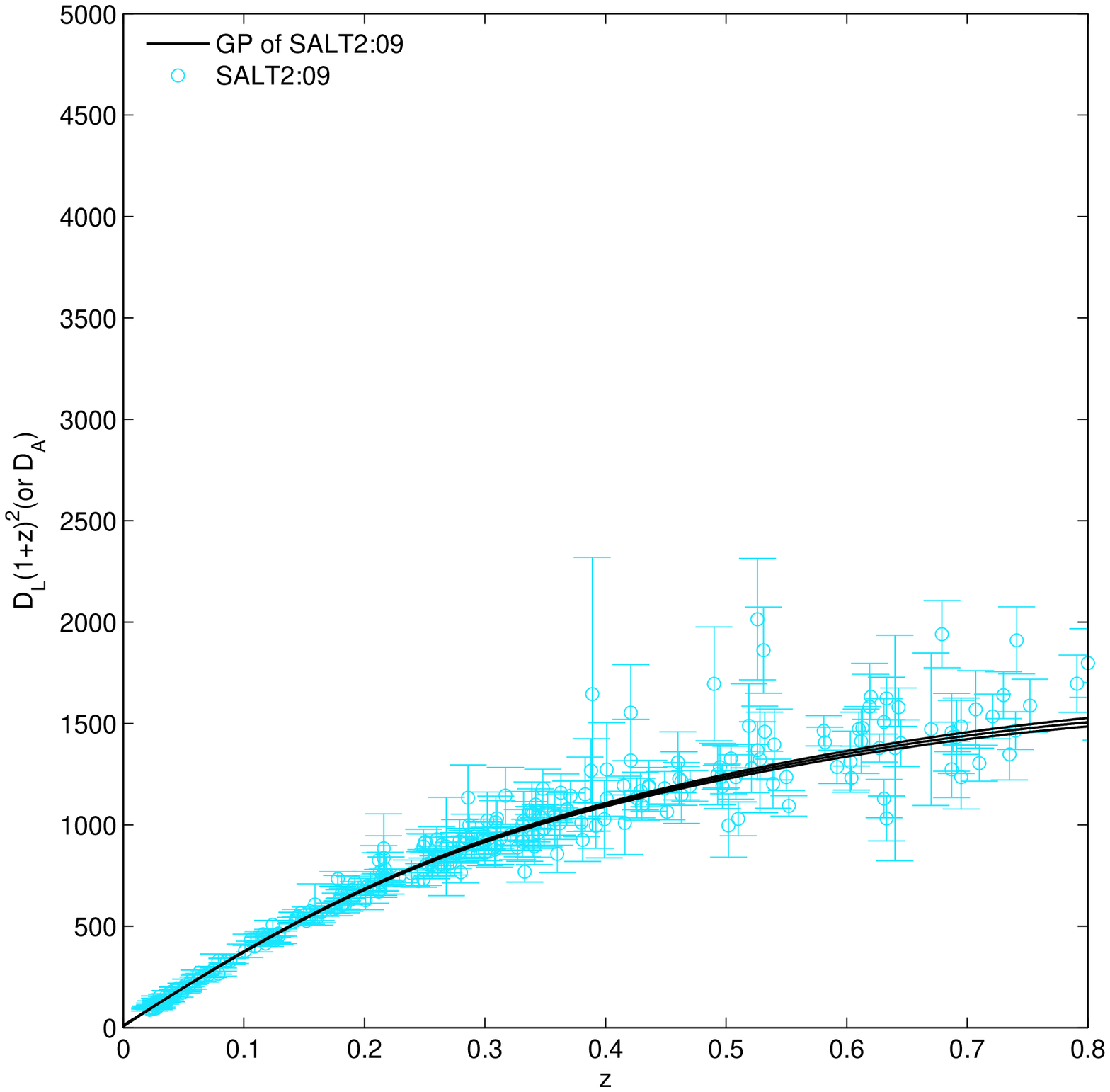}}\quad
\caption{ The upper panels show the GP result of the $D_{A}|_{cosm}$, ES$|_{ell}$ and SS$|_{circ}$ data which are obtained  from Ref.\cite{DeFilippis:2005hx}. The middle panels referred to the  GP of the SS data which are derived from Ref.\cite{Bonamente:2005ct}.  The lower panels present the GP of  LD data.}
  \label{gp}
\end{figure}


   \subsection{The GP of LD and ADD}
      In   Figure \ref{gp}, we present the   GP  results of the ADD  and LD data which show     the systematical  error of  GP result become smaller  compared to the  original data, especially in the dense data regime.  When two observational data  are very close,       the GP results favor the one with smaller error.  Thus, the GP results avoid large deviated data .

 Generally, the data from Ref.\cite{DeFilippis:2005hx} ($D_A|_{cosm}$, ES$|_{ell}$ and  SS$|_{circ}$) have the same  monotone increasing reconstructed shapes.   As  high redshift  data from Ref.\cite{DeFilippis:2005hx}  is rare,  it is reasonable  that the error of GP results become large as z increase.  The error of   GP result is  SS$|_{circ}>$ES$|_{ell}>D_A|_{cosm}$. 
  Specially,
the  GP result of $\eta$ shape of   SS$|_{circ}$   does not favor the $z=0.374$ (A370) and $z=0.784$ (MS 1137.5+6625) data.   

However,  the GP results of SS data (SS\Rmnum{1}, SS\Rmnum{2} and SS\Rmnum{3}) suffer from systematics of  heterogeneous data. The GP results seem  consistent with the observational data. But there is no explanation  for the high  $D_{A}$ value   around $z=0$. Because based on Eq.\ref{d},  the value of ADD data should be 0 when $z=0$. 
  This phenomenon is  caused by the lack of  SS data  in the low redshift.  To fix this problem, we use a  faked  data at $z=0$ with  $D_{A}=1 Mpc$ which is  close to the value derived   from the definition of standard ruler.   And its   errors $\sigma_{+}= 110 Mpc$,  $\sigma_{-}= 80 Mpc $ are from the lowest redshift data  in the SS sample where $z=0.14$. We call the SS with   fake data as  SS\Rmnum{1}$^{*}$, SS\Rmnum{2}$^{*}$ and SS\Rmnum{3}$^{*}$. 
Anyway,  the GP results of  SS\Rmnum{1}$^{*}$, SS\Rmnum{2}$^{*}$ and SS\Rmnum{3}$^{*}$  are oscillating.

  \subsection{The Comparisons}
  
\begin{figure}  \centering
      {\includegraphics[width=2.0in]{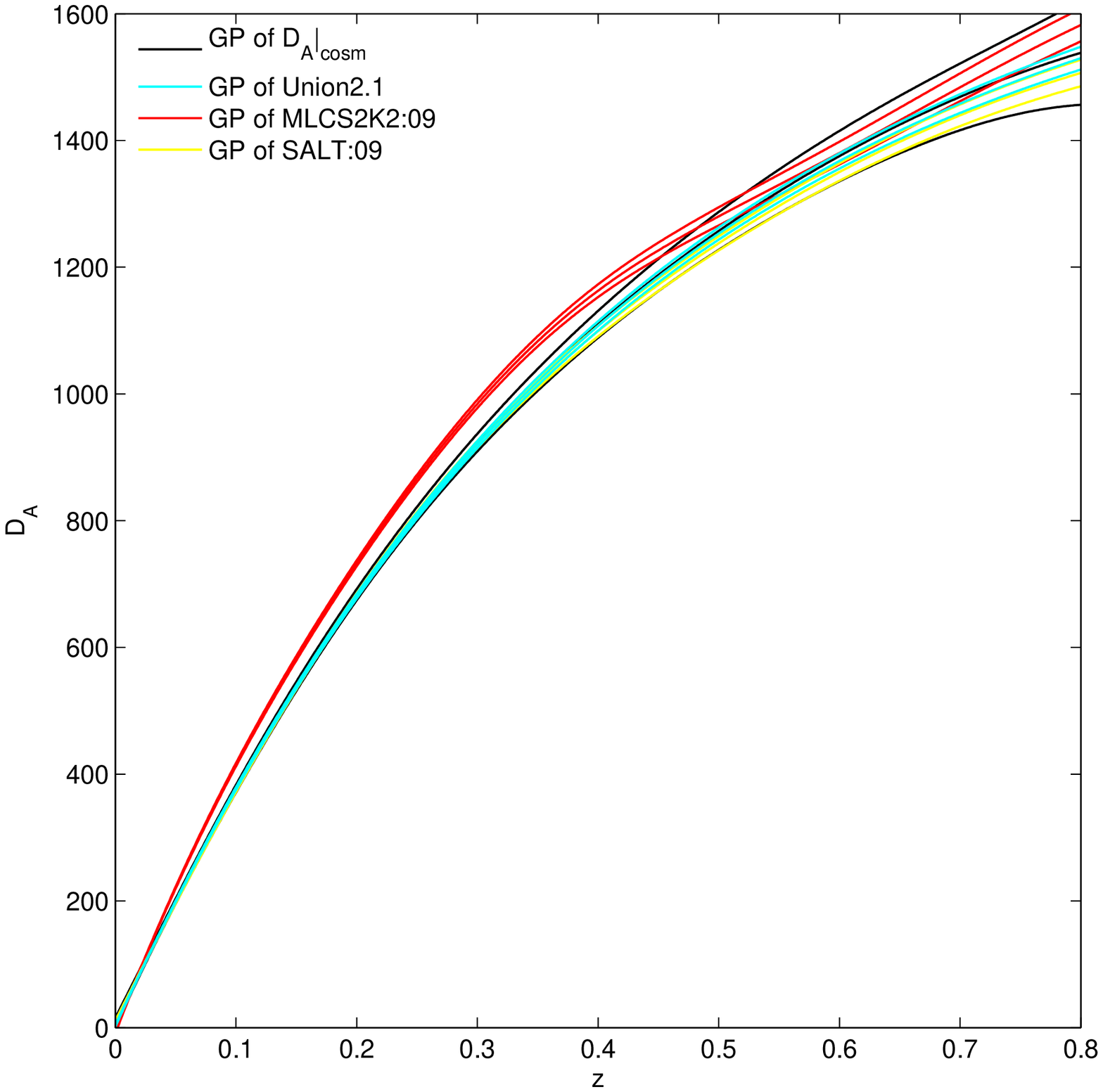}}\quad
       {\includegraphics[width=2.0in]{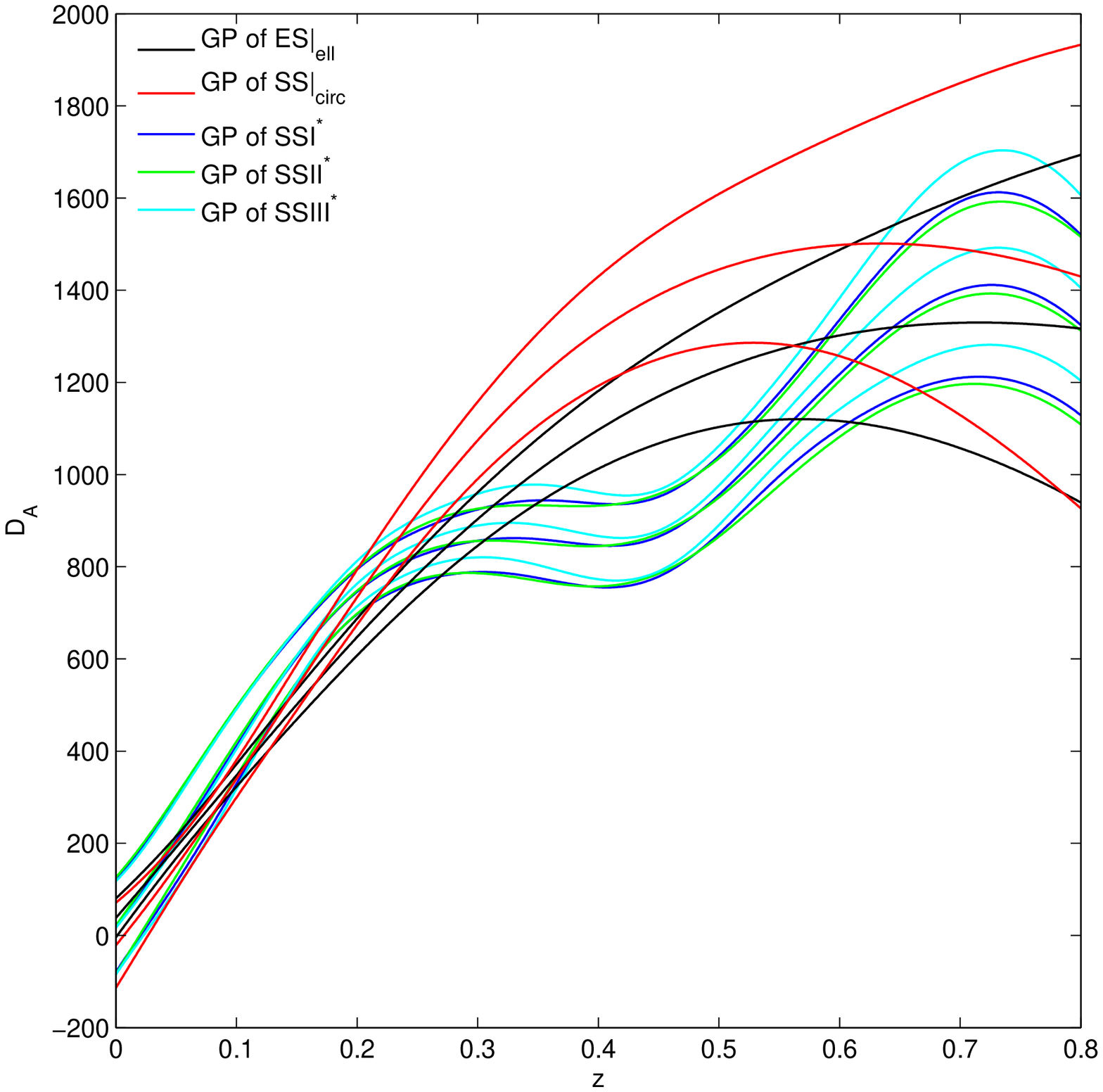}}\quad
       {\includegraphics[width=2.0in]{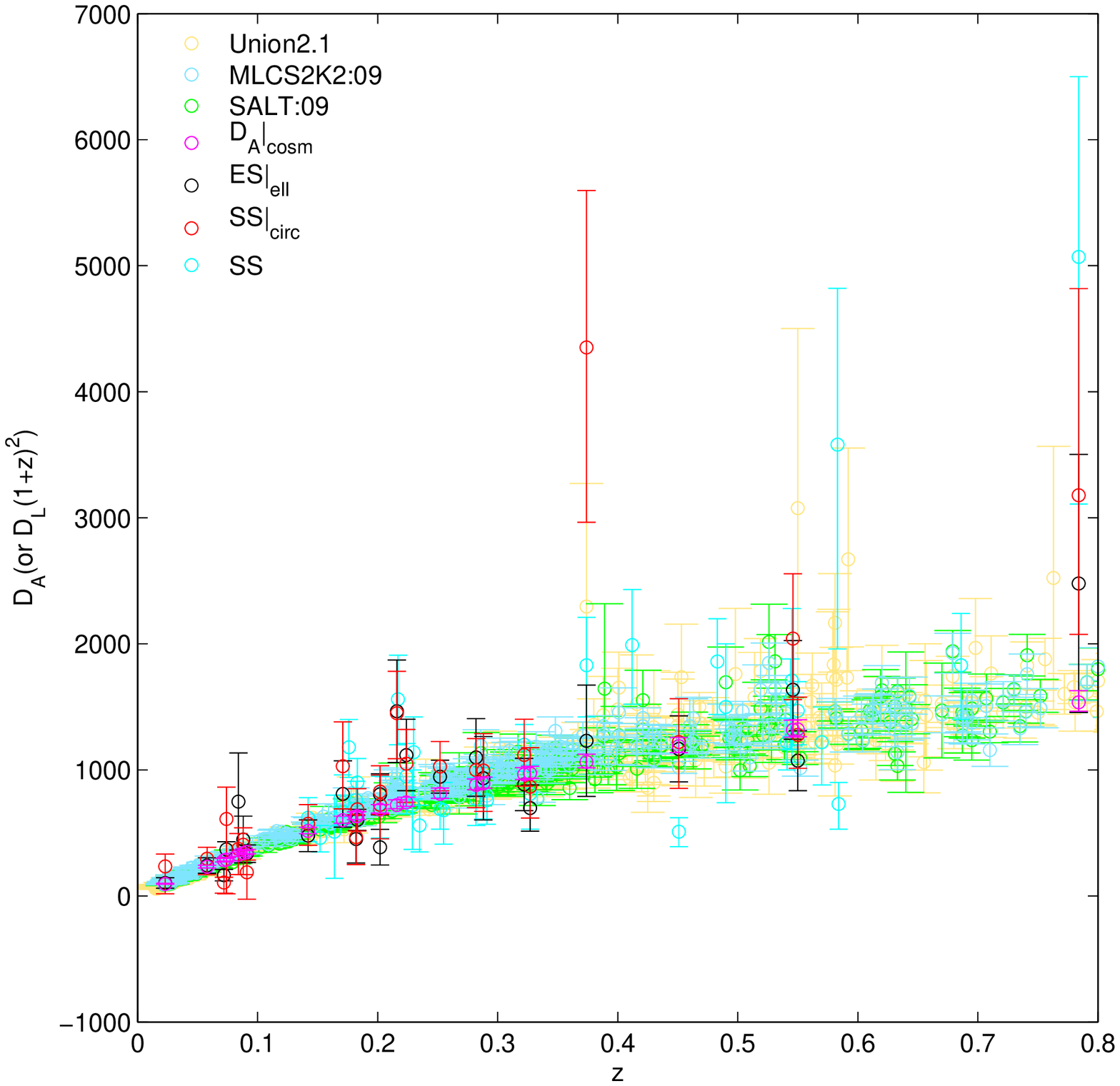}}\quad
\caption{The left panel presents the the GP results of the LD data and the $D_{A}|_{cosm}$ sample.  The middle panel shows the GP result of  the ES$|_{ell}$, SS$|_{circ}$ and  SS samples. The right panel shows the observational data. }
  \label{gpall}
\end{figure}

In the left panel of Figure \ref{gpall},   the $1\sigma$ region of the Union2.1, MLCS2K2:09  and SALT:09 are shown.     
 Roughly, the GP value of $D_{L}(1+z)^{2}$ is 
MLCS2K2:09 $>$ Union2.1 $\sim$ SALT:09.  The GP result of $D_A|_{cosm}$ nearly cover  all the GP of SNe  which indicates the $D_A|_{cosm}$  
    and LD data    are consistent.

As the middel panel of Figure \ref{gpall} shown,  the GP tendencies of ES$|_{ell}$  and SS$|_{circ}$ data are nearly the same in the low redshifts where the data   are concentrated. And because of the different galaxy morphology, the  ES$|_{ell}$  data give out  smaller GP result   at high redshift.
   It is expected  that the SS data (SS\Rmnum{1}, SS\Rmnum{2} and SS\Rmnum{3}) should have the same GP tendency with 
 the SS$|_{circ}$ data because both of them are assumed spherical symmetry. Anyway, the SS  obtain a oscillating behavior while the SS$|_{circ}$ data  obtain a monotone increasing function.
Because of different error dealing methods, the GP result show SS \Rmnum{3} (GP) $>$ SS \Rmnum{1}  (GP) $\sim$ SS \Rmnum{2} (GP) with slight differences.  The GP of SS \Rmnum{3}  seems closest to  the LD data.  
 The oscillating behavior may be caused by GP but it does not affect the GP of $\eta$ as we will see.

\section{Reconstruction of DD relation}\label{sec5}

\subsection{$\eta$ with One GP}


\begin{figure}  \centering
    {\includegraphics[width=2.0in]{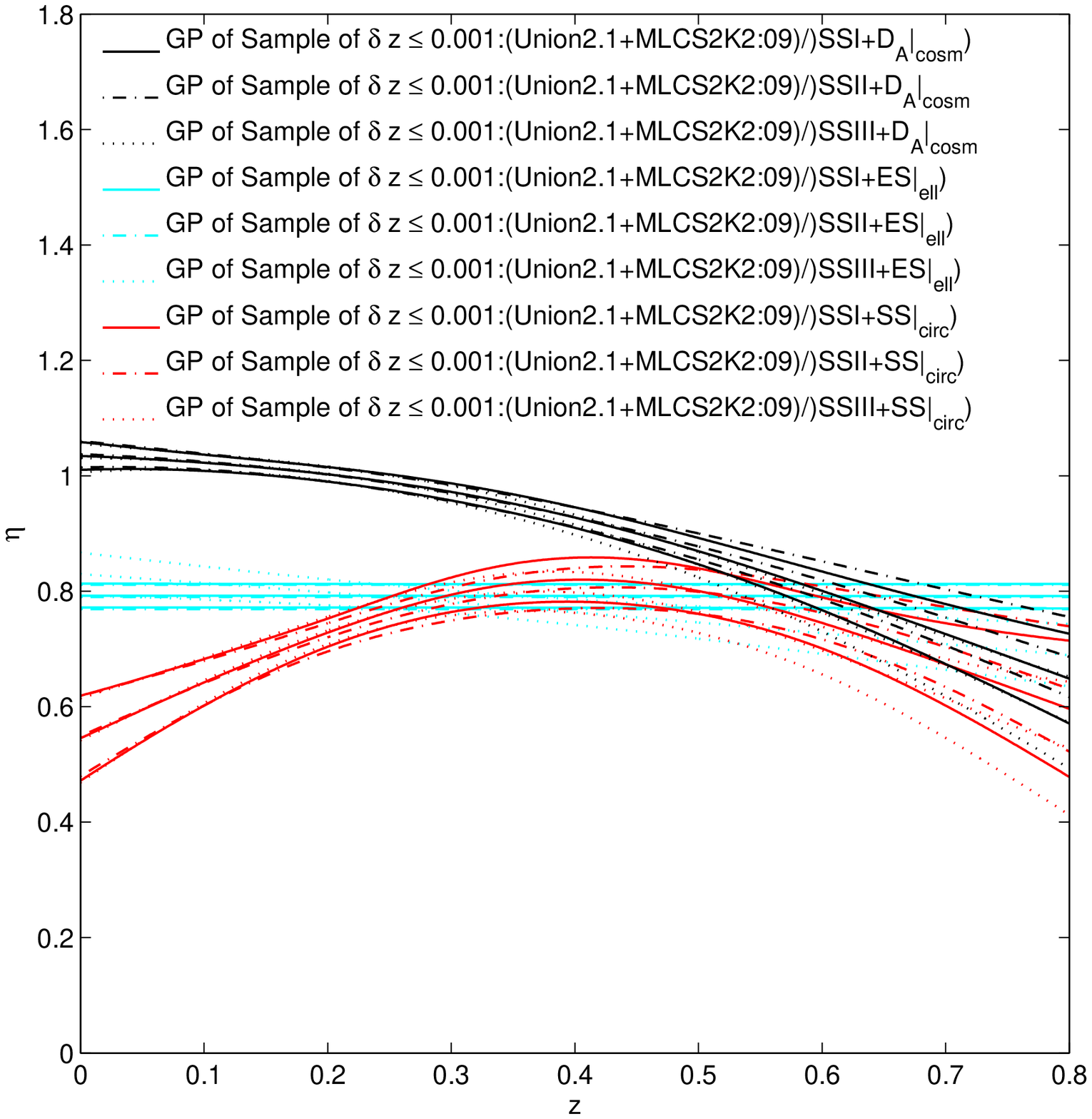}}\quad
        {\includegraphics[width=2.0in]{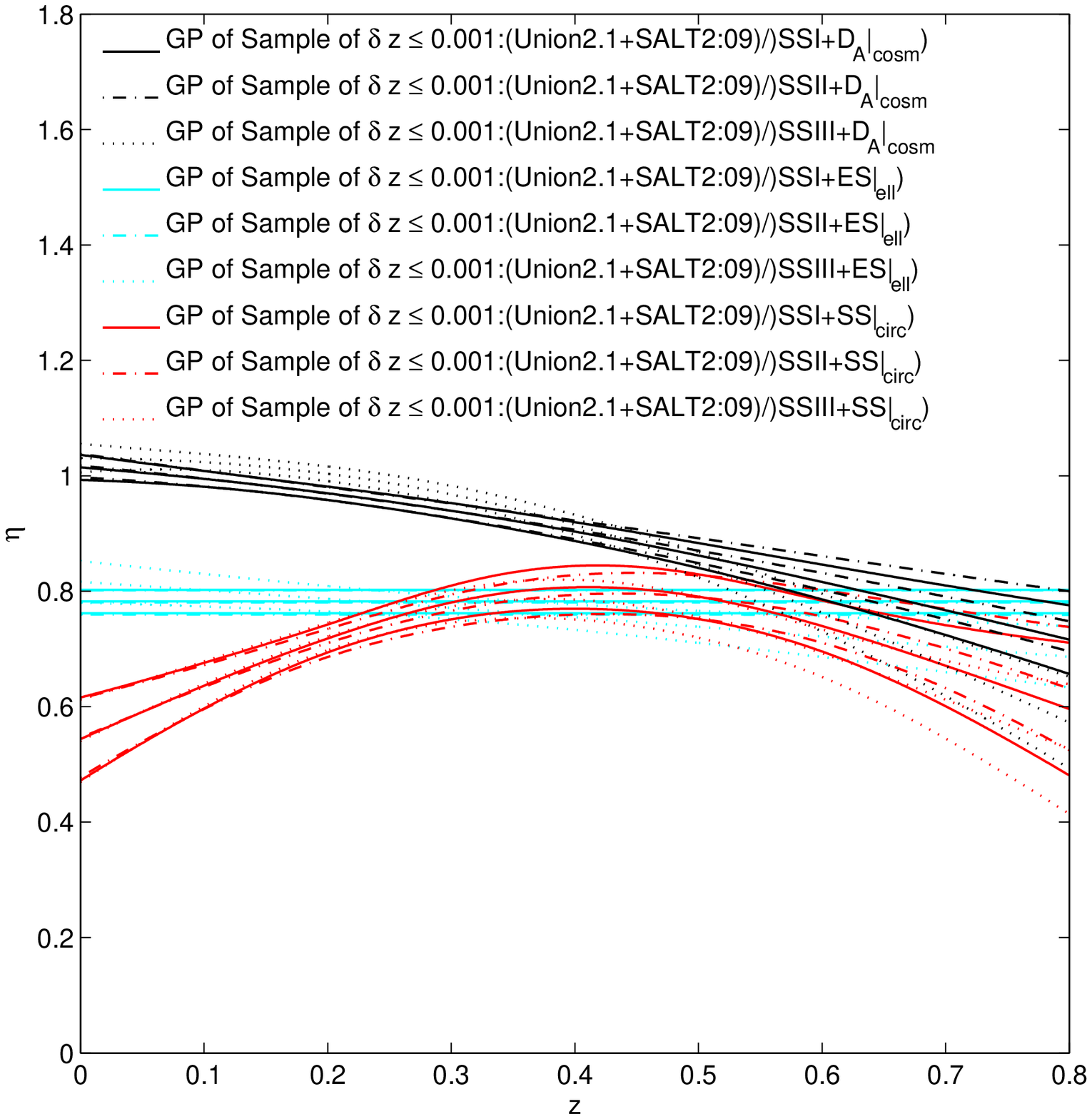}}\quad
\caption{ 
The panels are for the overlap samples of   $\delta z\leq 0.001$.}
  \label{0001all}
\end{figure}

\begin{figure}  \centering
  {\includegraphics[width=2.0in]{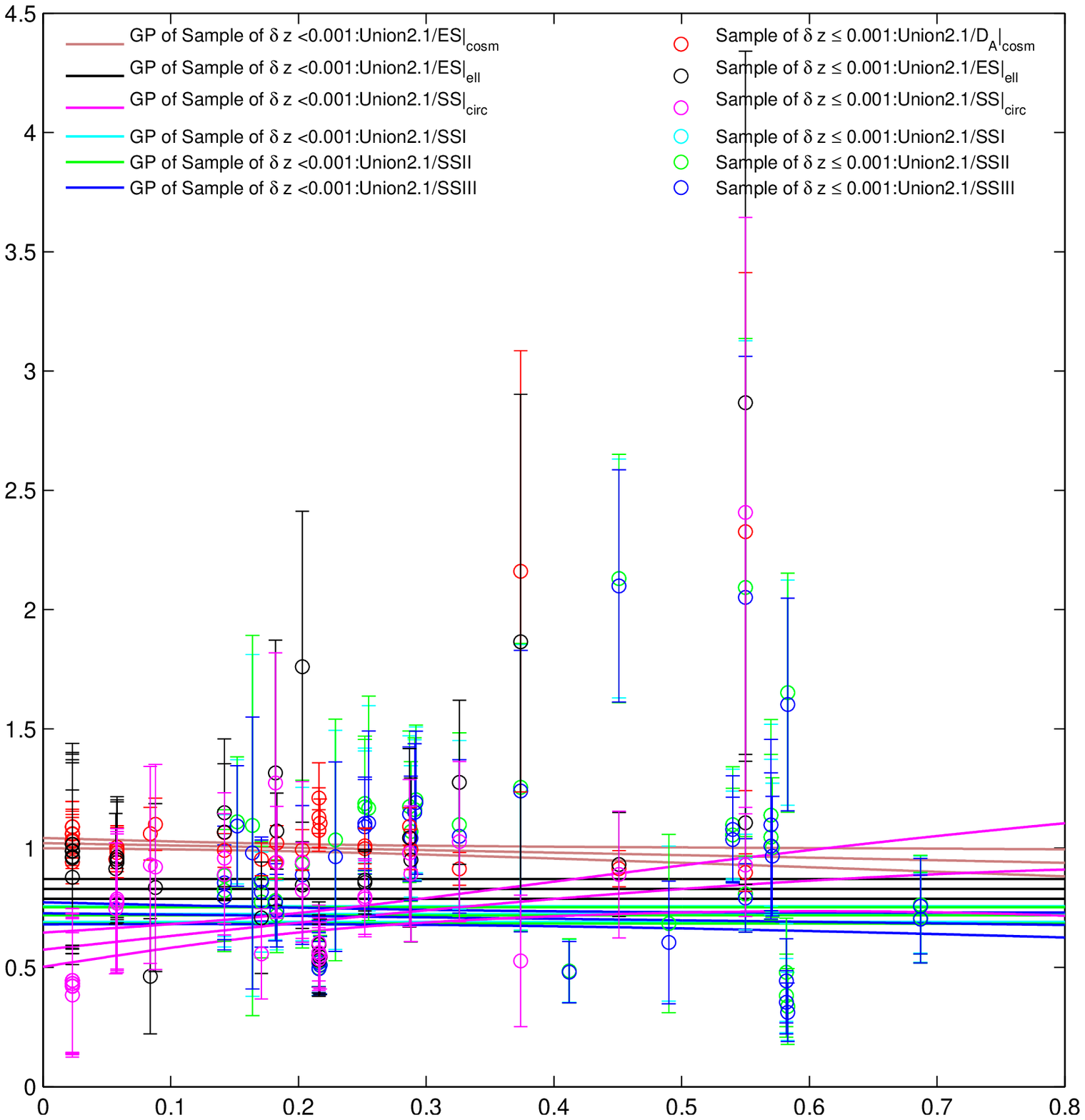}}\quad
    {\includegraphics[width=2.0in]{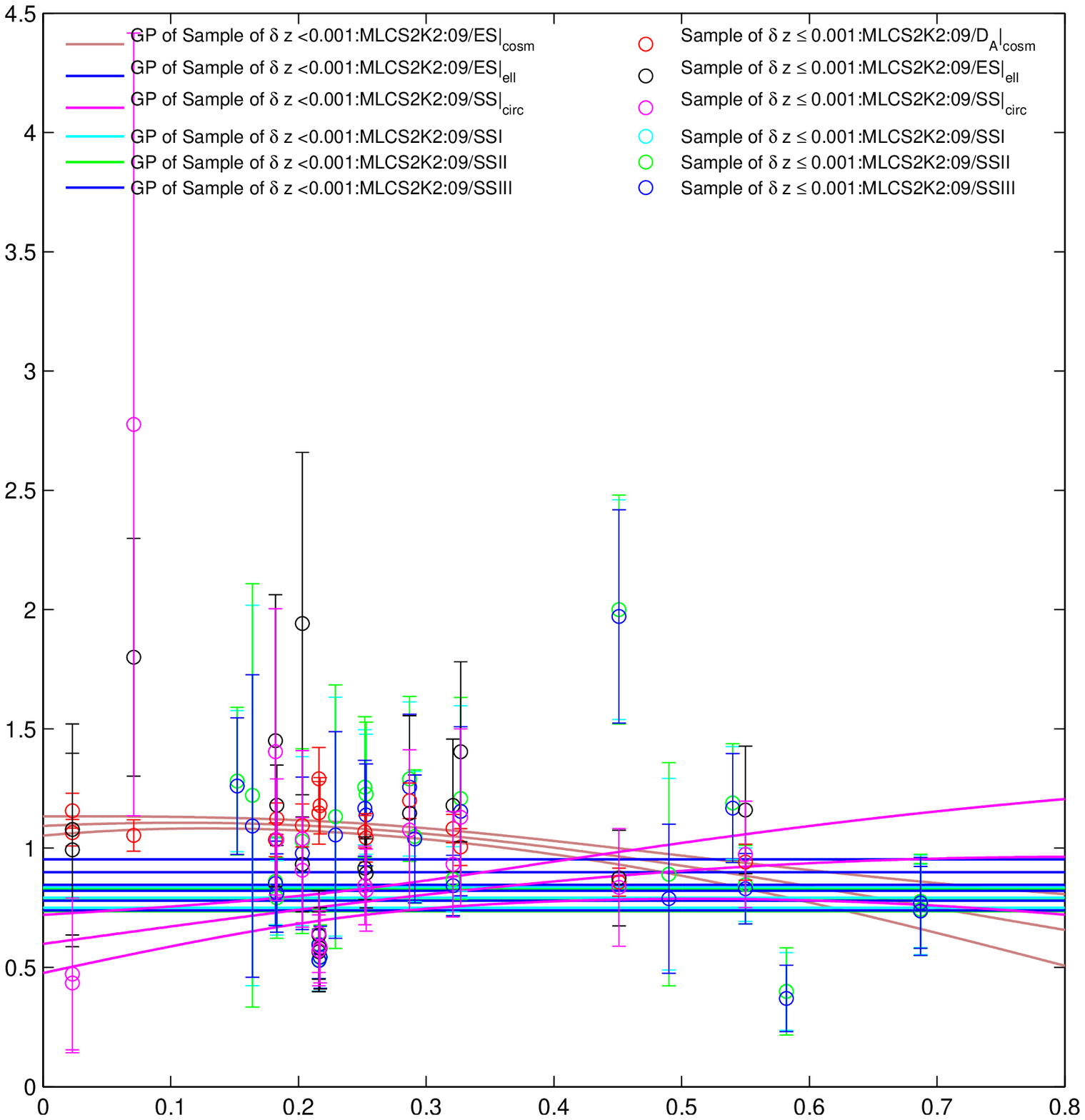}}\quad
      {\includegraphics[width=2.0in]{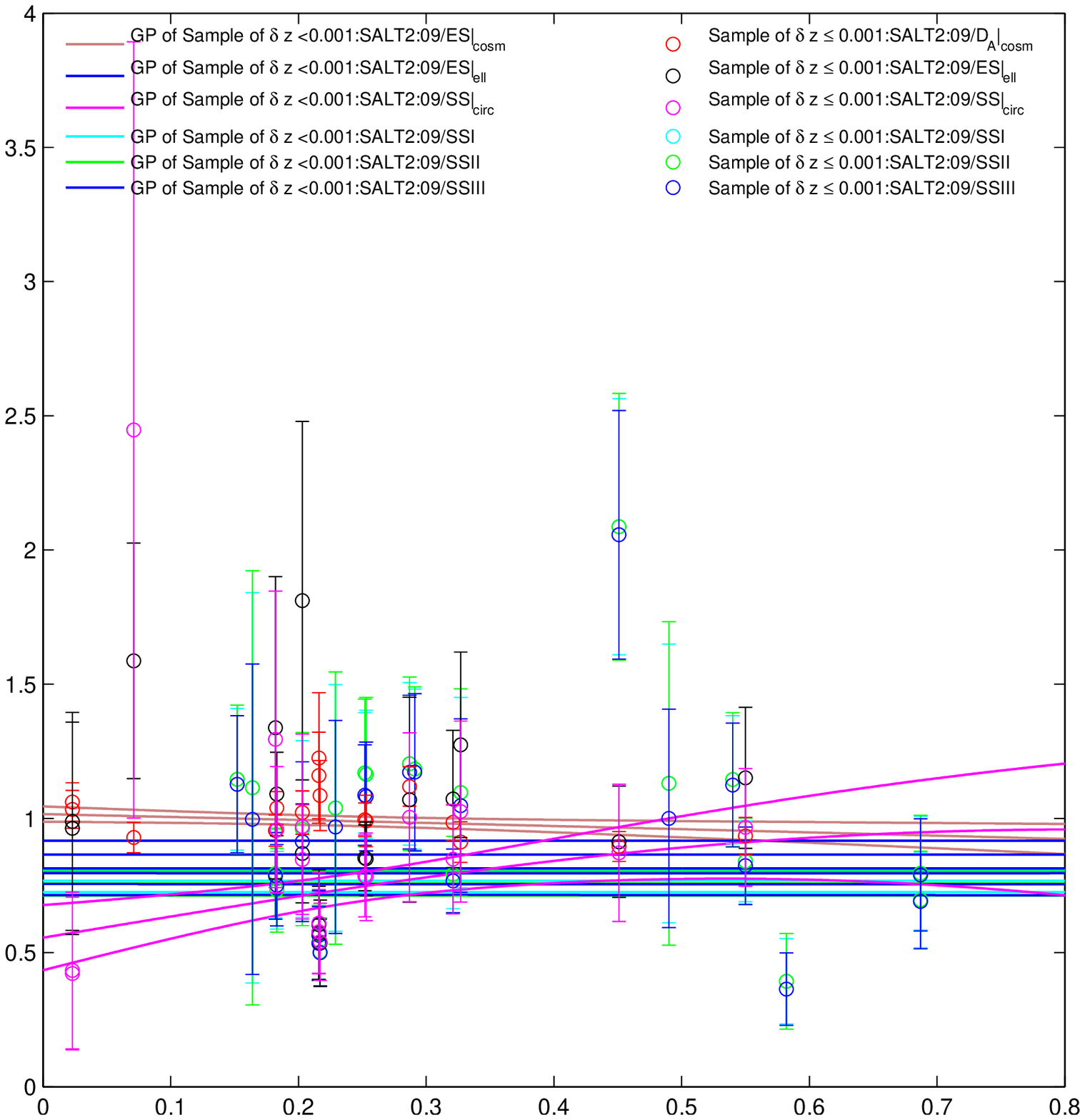}}\quad
\caption{ The  panels are for the separate samples of $\delta z\leq 0.001$.}
  \label{0001}
\end{figure}


 \begin{table*}[t]
 \tiny
\begin{center}
\begin{tabular}{|ll|ll|ll|}
\hline $\eta(GP)$ & $\eta_0$ & $\eta(GP)$ & $\eta_0$ & $\eta(GP)$&  $\eta_0$\\
\hline Union2.1/SS \Rmnum{1} &$0.72\pm  0.03$ &MLCS2K2:09/SS \Rmnum{1} & $0.79\pm  0.04$ & SALT2:09/SS \Rmnum{1} & $  0.77\pm  0.04$\\
 Union2.1/SS \Rmnum{2} &$ 0.72\pm  0.03$&MLCS2K2:09/SS \Rmnum{2}  &$0.78\pm  0.05$ & SALT2:09/SS \Rmnum{2} & $  0.76\pm  0.04$\\
 Union2.1/SS \Rmnum{3} &$0.73\pm  0.05$ &MLCS2K2:09/SS \Rmnum{3}  &$0.78\pm  0.04$ & SALT2:09/SS \Rmnum{3} & $ 0.76\pm  0.04$\\
 Union2.1/$D_A|_{cosm}$  &$1.02\pm  0.02$ & MLCS2K2:09/$D_A|_{cosm}$   &$ 1.09\pm  0.04$& SALT2:09/$D_A|_{cosm}$  &$1.02\pm  0.03$\\
 Union2.1/ES$|_{ell}$  &$0.83\pm  0.04$ & MLCS2K2:09/ES$|_{ell}$   &$ 0.90\pm  0.05$& SALT2:09/ES$|_{ell}$  &$0.87\pm  0.05$\\ 
 Union2.1/SS$|_{circ}$  &$0.57\pm  0.07$ & MLCS2K2:09/SS$|_{circ}$  &$0.60\pm  0.12$& SALT2:09/SS$|_{circ}$  &$0.56\pm  0.12$\\
\hline Union2.1/$D_A|_{cosm}$(GP) &$0.99\pm  0.01$ &MLCS2K2:09/$D_A|_{cosm}$(GP) & $1.07\pm  0.01 $ & SALT2:09/$D_A|_{cosm}$ (GP) & $ 0.98\pm  0.01 $\\
 Union2.1/ES$|_{ell}$(GP)&$0.84\pm  0.04 $&MLCS2K2:09/ES$|_{ell}$(GP) &$ 0.90\pm  0.05$ & SALT2:09/ES$|_{ell}$ (GP)& $  0.83\pm  0.04 $ \\
 Union2.1/SS$|_{circ}$(GP)  &$0.64\pm  0.07$ &MLCS2K2:09/SS$|_{circ}$(GP)&$  0.70\pm  0.08$ & SALT2:09/SS$|_{circ}$ (GP) & $ 0.64\pm  0.07$\\
 \hline Union2.1(GP)/$D_A|_{cosm}$  &$1.01\pm  0.01$ & MLCS2K2:09(GP)/$D_A|_{cosm}$   &$ 1.11\pm  0.01$& SALT2:09(GP)/$D_A|_{cosm}$  &$ 1.01\pm  0.01$\\
 Union2.1(GP)/ES$|_{ell}$   &$0.89\pm  0.03$ & MLCS2K2:09(GP)/ES$|_{ell}$  &$1.17\pm  0.02  $& SALT2:09(GP)/ES$|_{ell}$  &$1.05\pm  0.03$\\ 
 Union2.1(GP)/SS$|_{circ}$  &$1.46\pm  0.07$ & MLCS2K2:09(GP)/SS$|_{circ}$   &$ 1.47\pm  0.08$& SALT2:09(GP)/SS$|_{circ}$  &$1.34\pm  0.07$\\
\hline
\end{tabular}
\end{center}
\caption[crit]{\small The  GP values of $\eta_0$ for  selected $\eta$ are listed.   The left, middle and right columns correspond to Union2.1, MLCS2K2:09 and SALT2:09 related data respectively. }
\label{tab2}
\end{table*}

The model-independent criterion that the redshift difference of 
the ADD and LD data should be less than $0.005$ was used. 
In order to avoid the appearance of too many $\eta$ data for one redshift,  the criterion is improved to  $\delta z = |z_{LD}-z_{ADD} |\leq 0.001$ in this letter. Thus we have the so-called  $\delta z \leq 0.001$   sample by combining all the uncertainties in quadrature.
The derived  values of $\eta$ are more centered around the $\delta z=0 $ line compared to Refs.\cite{Li:2011exa,Holanda:2010vb}.  The  redshift   range of the SS related sample  is $0.142<z<0.686$, while that of   the ES related sample  is $0.023<z<0.550$.  The exact number of  $\delta z \leq 0.001$ sample at each ADD redshift are summarized in Table \ref{tab}. In principle,  the GP could be applied  to see the tendency of $\eta$ instead of assuming a parameterization of $\eta$.

\subsubsection{Overlap Sample}
In Figure \ref{0001all},  the Union2.1
and MLCS2K2:09 (or SALT2:09) data are combined as  LD data. 
Meanwhile,  the SS (SS\Rmnum{1}, SS\Rmnum{2} or SS\Rmnum{3}) data from Ref.\cite{Bonamente:2005ct} and the samples  from Ref.\cite{DeFilippis:2005hx} ($D_{A}|_{cosm}$, or
 ES$|_{ell}$, or SS$|_{circ}$) are combined as  ADD data. The $\eta$ value of the Union2.1+ MLCS2K2:09/ADD sample  seems  smaller than that of  the Union2.1+SALT2:09/ADD sample.
 Though the SS\Rmnum{1}, SS\Rmnum{2}, SS\Rmnum{3} related data are varied with the  error dealing methods,  the  error dealing methods just  affect   their reconstruction of  $\eta$ on the value slightly. 
And, only the ES$|_{ell}$ related data give out  constant $\eta$ results  which is around $\eta_0=0.8$ and not consistent with  the DD relation.
The $\eta$  samples which are related to spherical symmetry (SS+SS$|_{circ}$) shape like  bulge.     
 The $D_A|_{cosm}$ related data  are supposed to have a $\eta$ constant reconstruction. But  it seems to be affected by the galaxy cluster sample of spherical symmetry   which makes a declining trend of $\eta$.  
Based on the above discussions, we do not regard the combination of the SS and the ES models  a good idea.

\subsubsection{ Separated Samples}
In this section, we will discuss the $\eta$ sample derived from the separated LD and  ADD data. 
 The number  of every separated  sample   is enough to have an effective GP as Figure \ref{0001} shows.
As for the spherical sysmmetry of cluster,   the SS$|_{circ}$ related data give out a monotone increasing function. However,
all the other samples give out a nearly constant $\eta$. The MLCS2K2:09/ES$|_{ell}$  data has $\eta=0.90\pm0.05$  which is closest to $\eta=1$.  And, there is no physical explanation for the difference between the   SS\Rmnum{1} ( or SS\Rmnum{2} or SS\Rmnum{3})    and   SS$|_{circ}$ related data.
Respect to  the reconstructing results of Union2.1/$D_A|_{cosm}$ and SALT2:09/$D_A|_{cosm}$ in Figure \ref{0001}  and Table \ref{tab2}, the 
DD relation($\eta=1$)  is faved.    Meanwhile,  the MLCS2K2:09/$D_A|_{cosm}$  result  gives out a decreasing tendency and gets $\eta_0= 1.09\pm  0.04$  with  a departure  from the DD relation.  
 The difference between the MLCS2K2 and SALT2 fitter will appear in the following.

\subsection{ GP with  Two GPs }
    \begin{figure}  \centering
       {\includegraphics[width=2.0in]{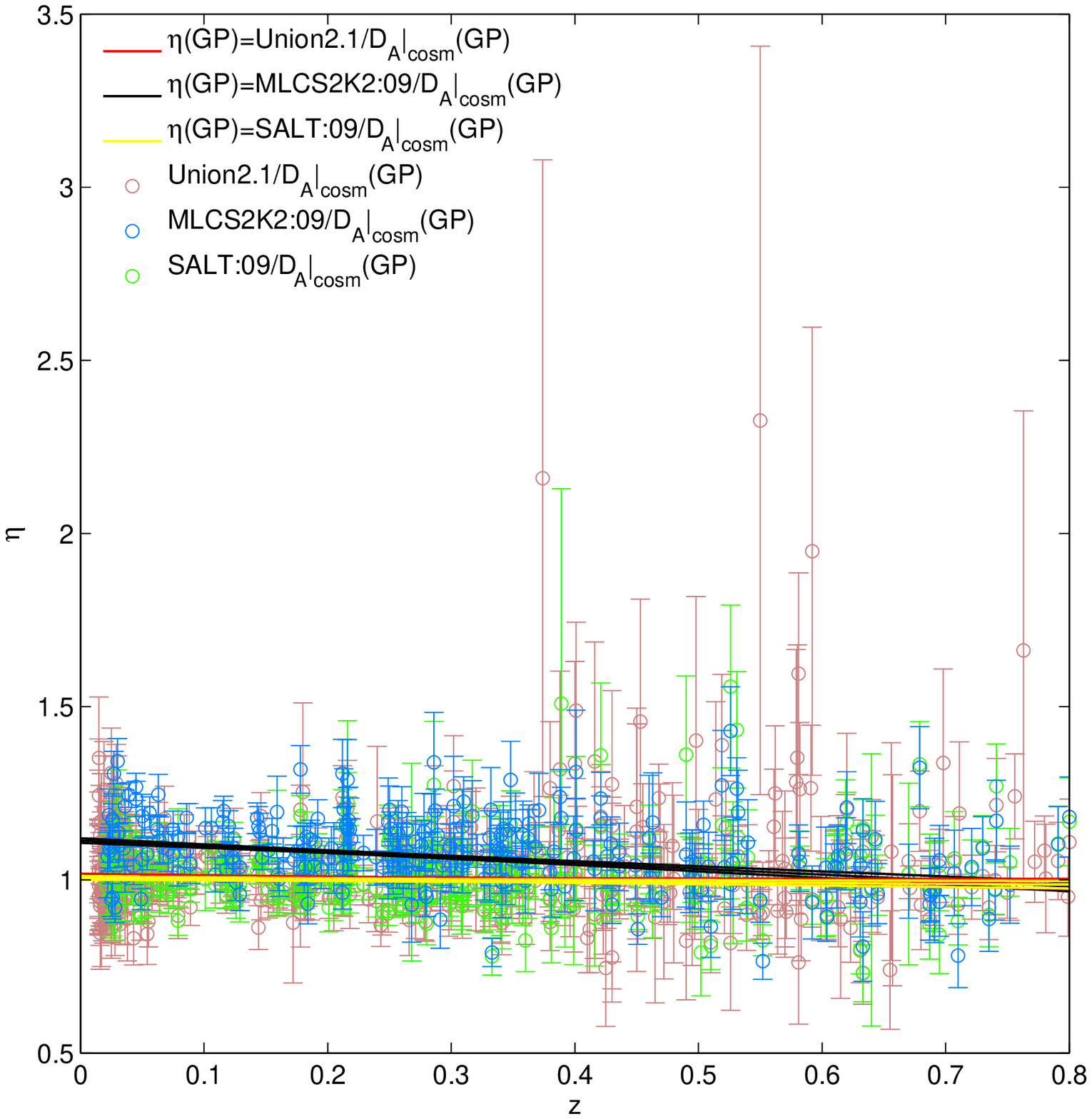}}\quad
   {\includegraphics[width=2.0in]{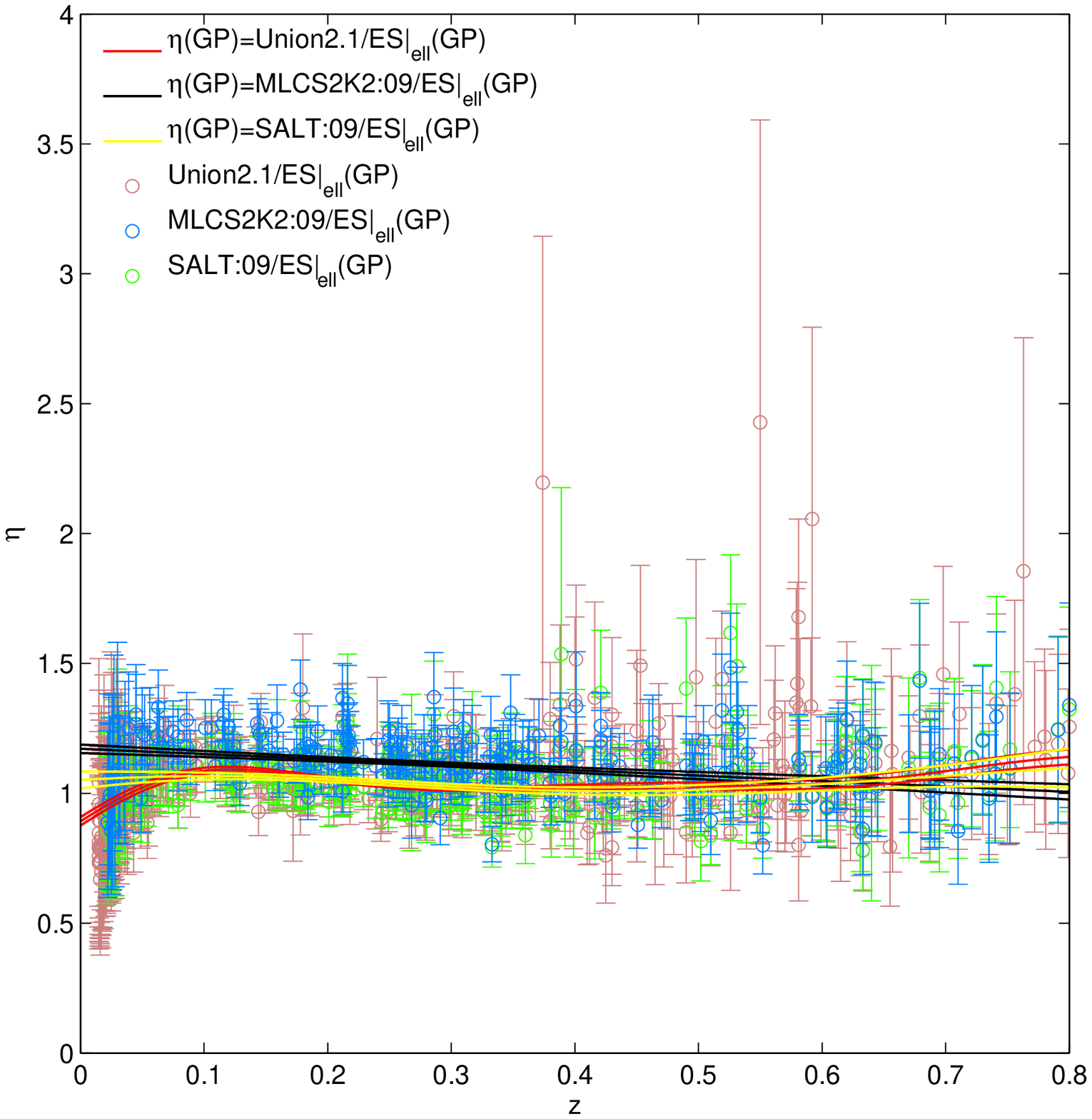}}\quad
   {\includegraphics[width=2.0in]{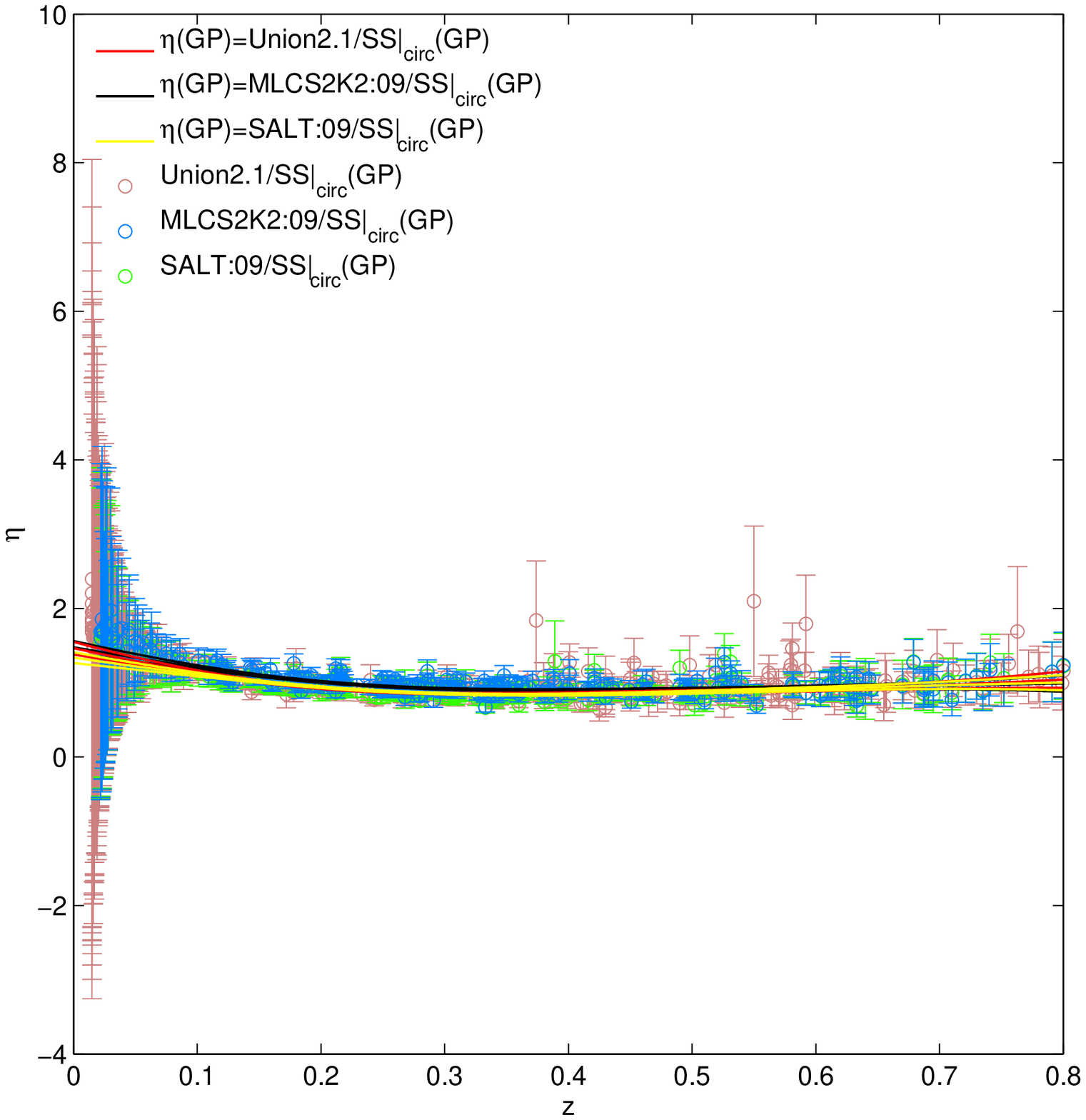}}\quad\\
              {\includegraphics[width=2.0in]{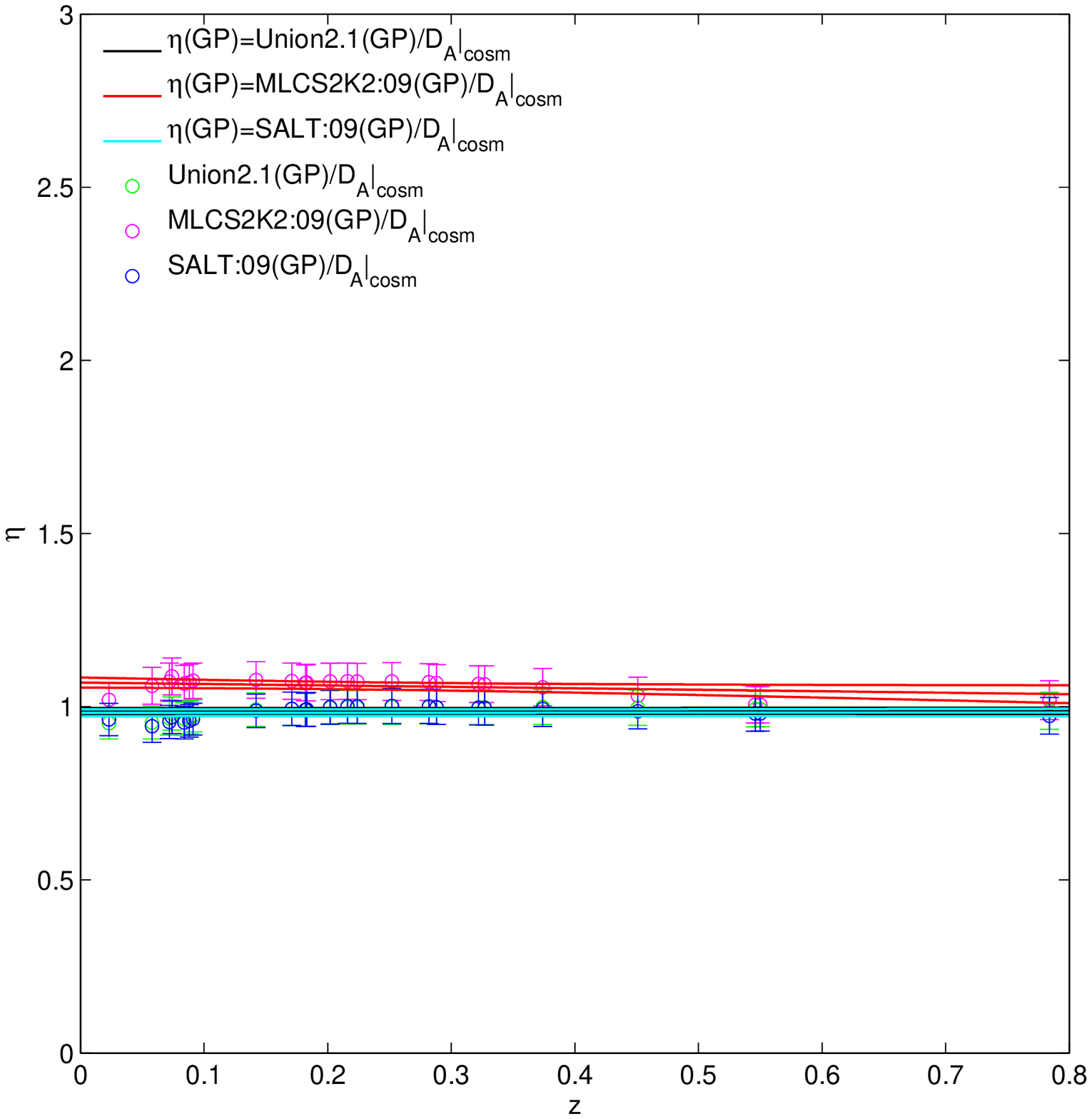}}\quad 
  {\includegraphics[width=2.0in]{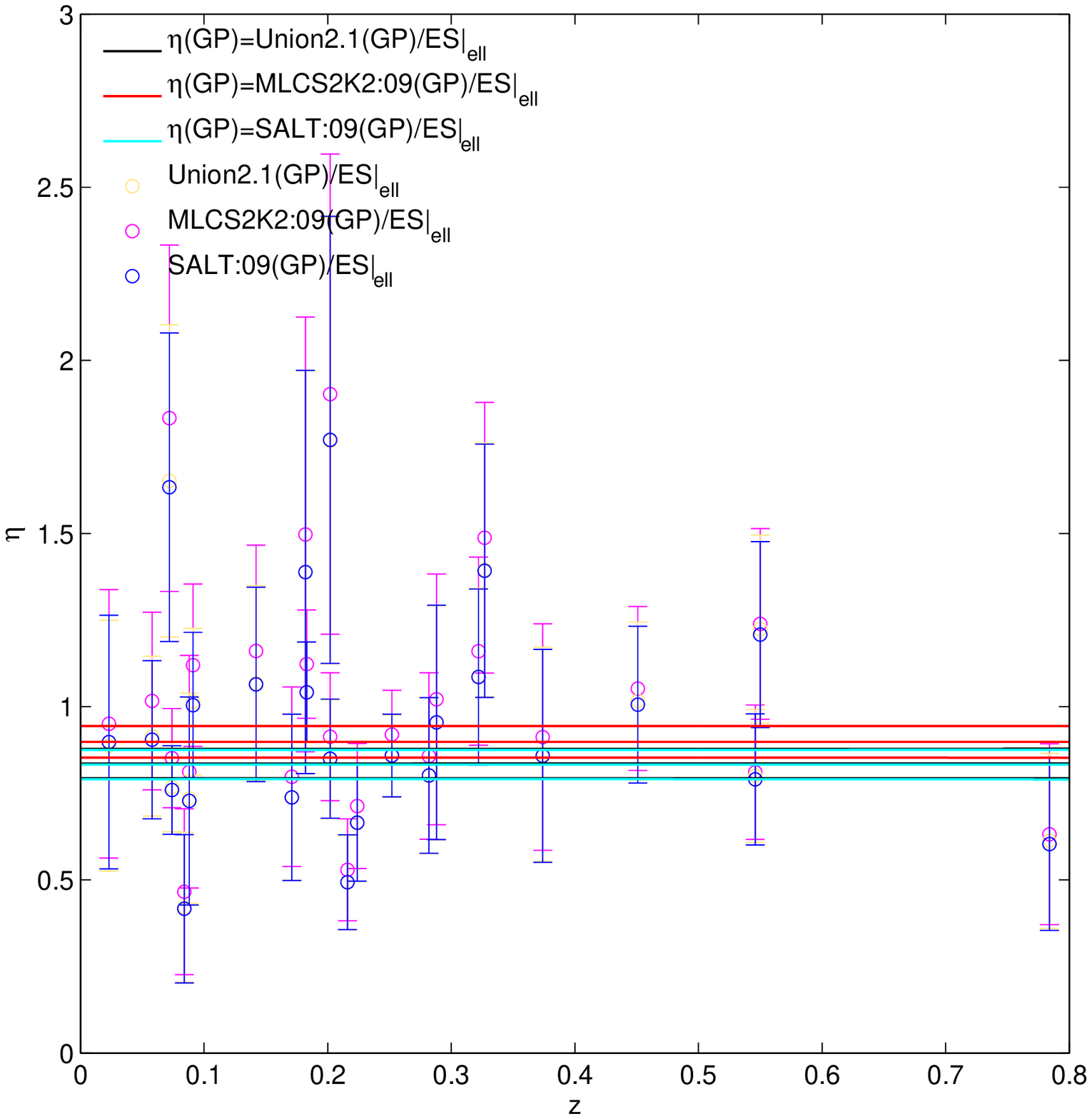}}\quad
    {\includegraphics[width=2.0in]{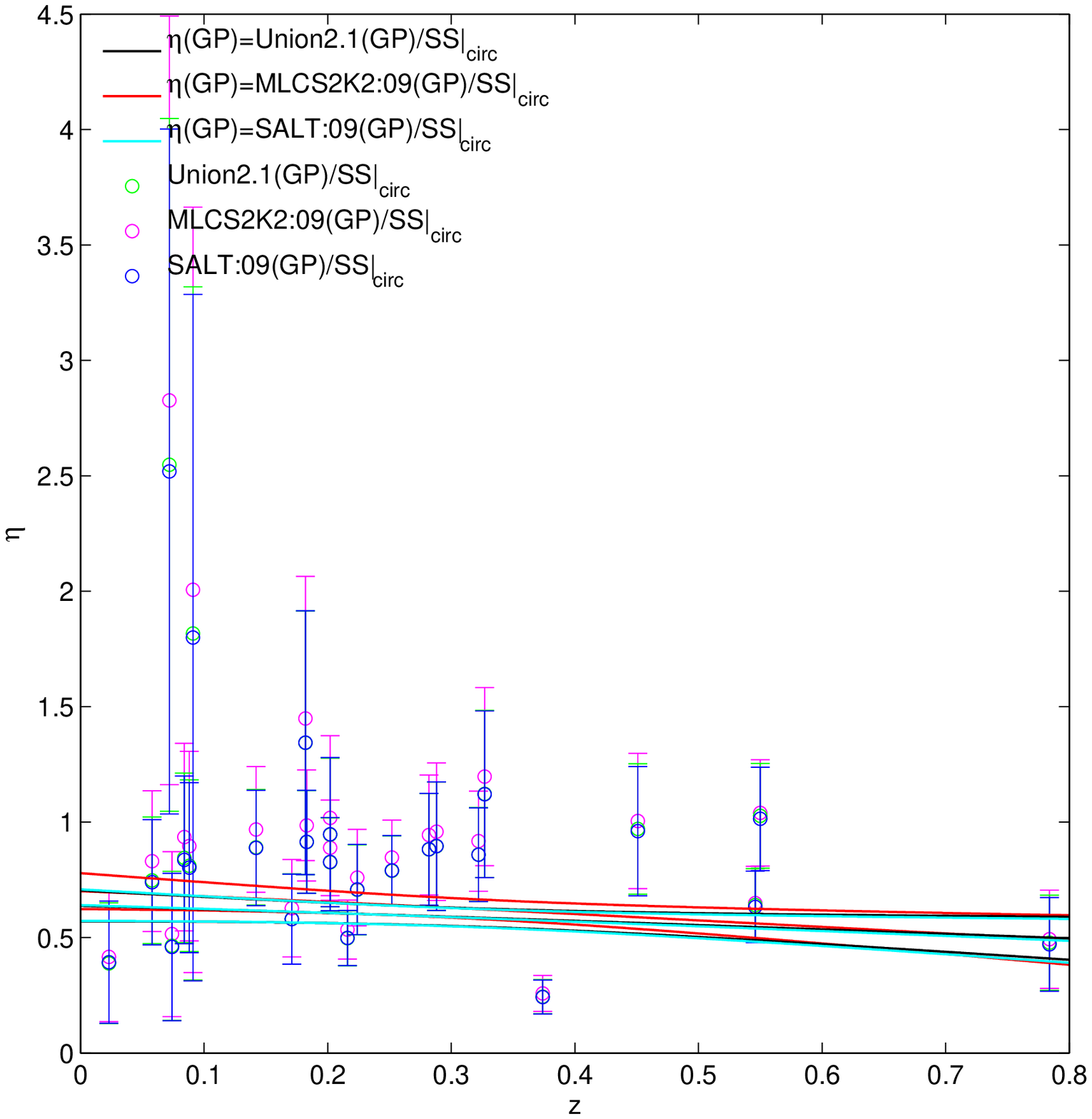}}\quad\\
     {\includegraphics[width=2.0in]{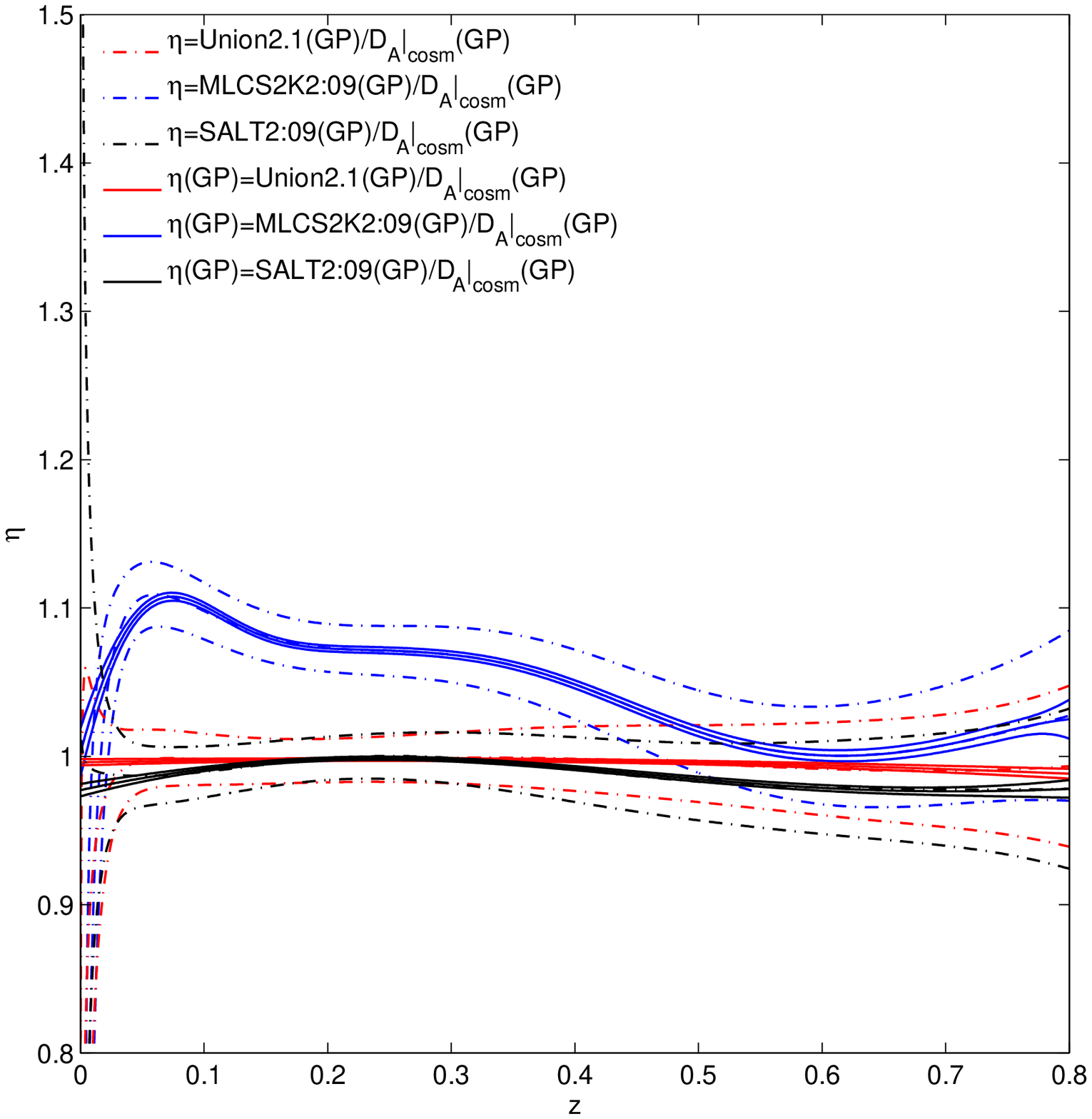}}\quad
      {\includegraphics[width=2.0in]{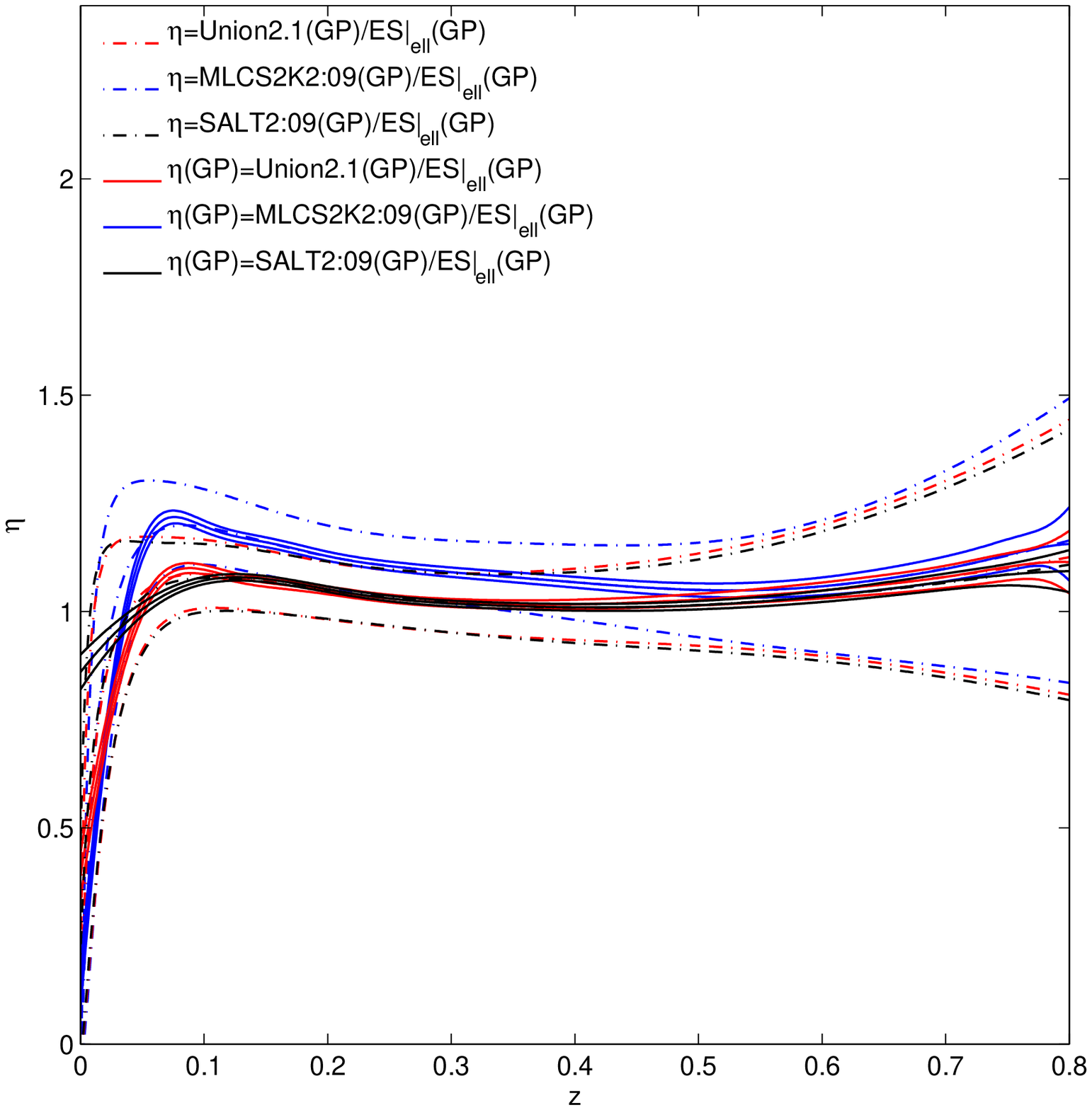}}\quad
    {\includegraphics[width=2.0in]{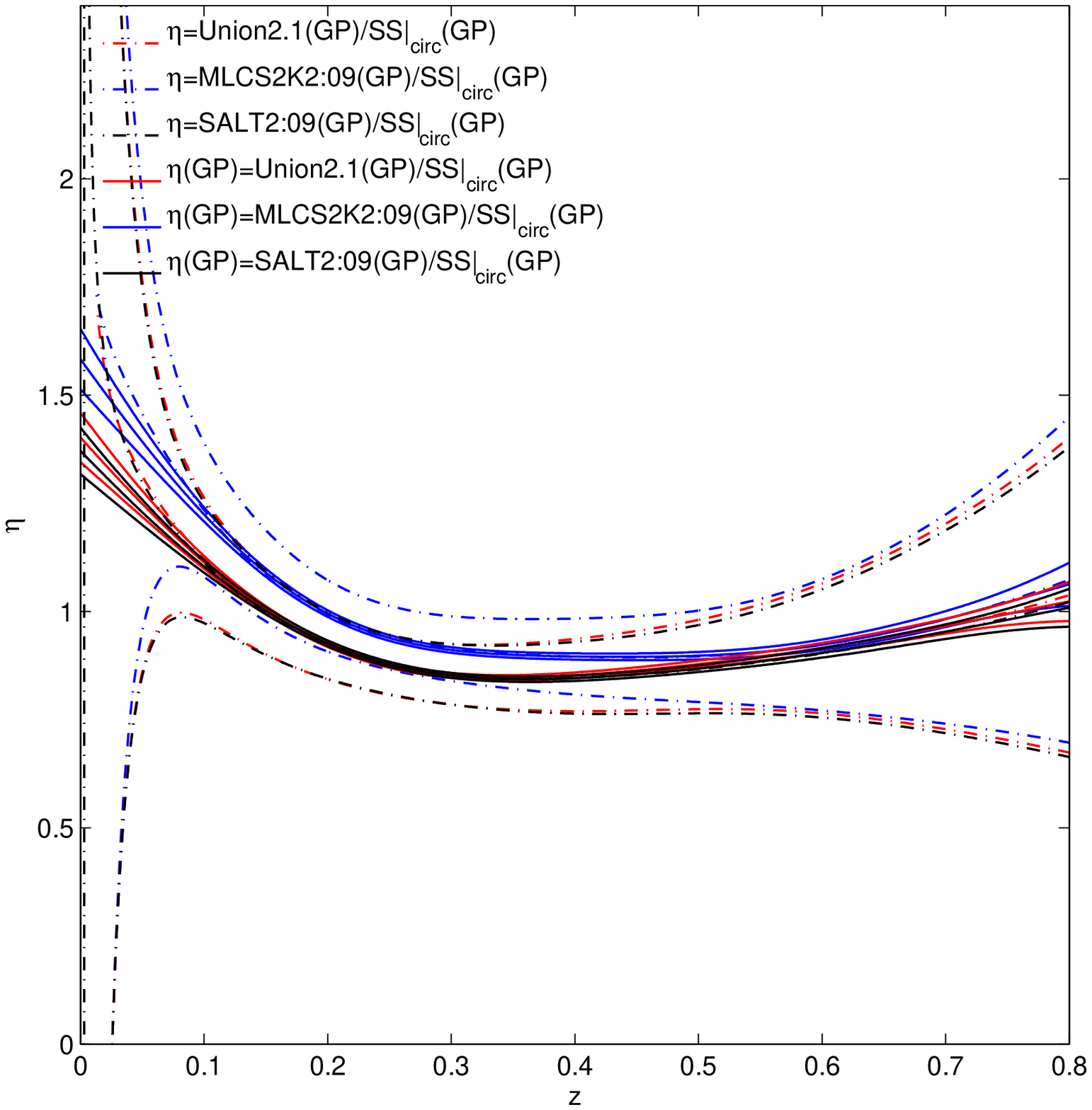}}\quad
\caption{The upper, middle and lower  panels are  the GP results of $\eta$ of LD(GP)/ADD, LD/ADD(GP) and LD(GP)/ADD(GP) respectively where the ADD data are derived from Ref.\cite{DeFilippis:2005hx}. The points    represent the observational data  with its error. The dashed lines are for $\eta= LD(GP)/ADD(GP)$. The solid lines are for the GP results.  }
  \label{2gpall}
\end{figure}

\begin{figure}  \centering
 {\includegraphics[width=2.0in]{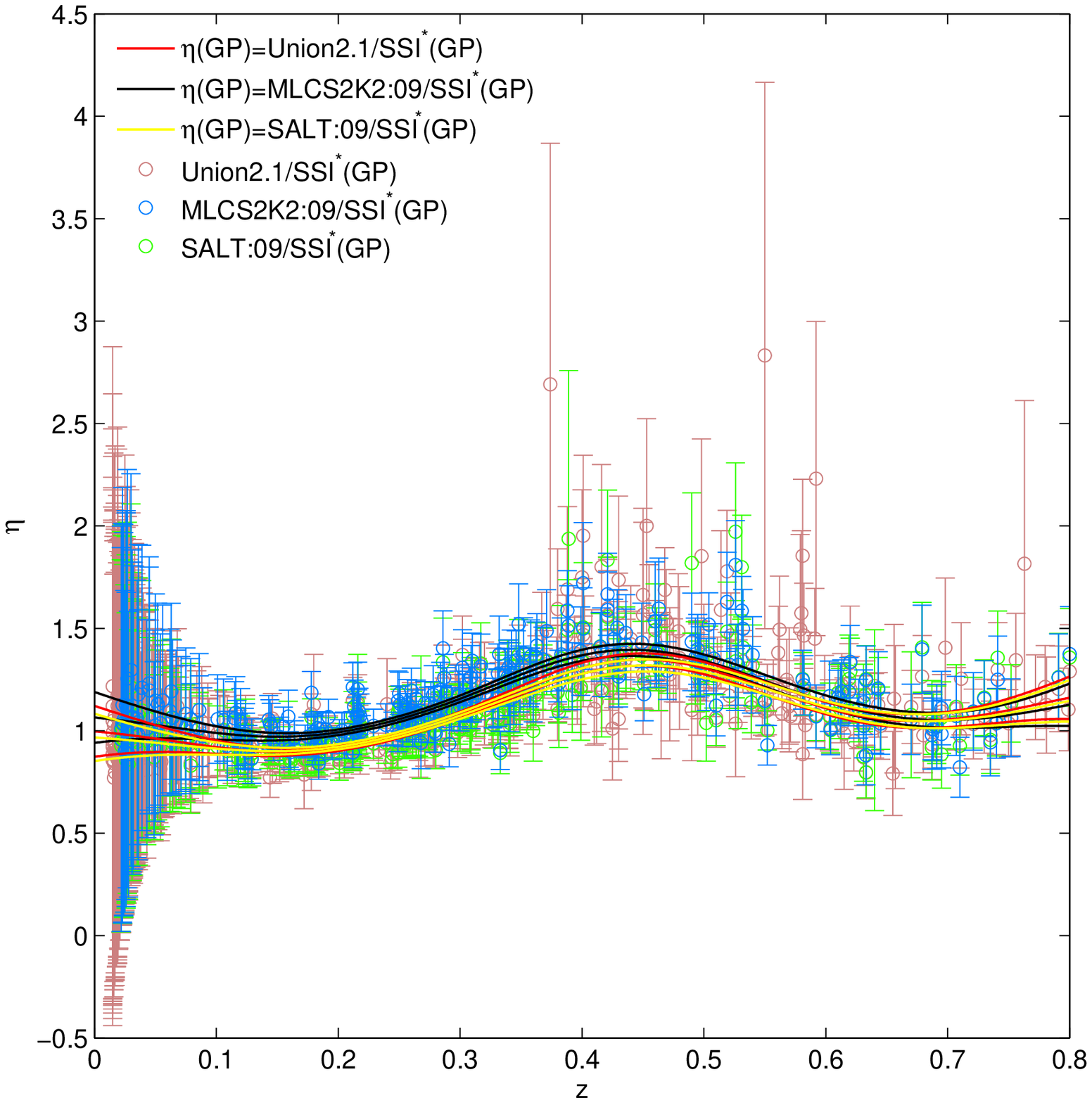}}\quad
    {\includegraphics[width=2.0in]{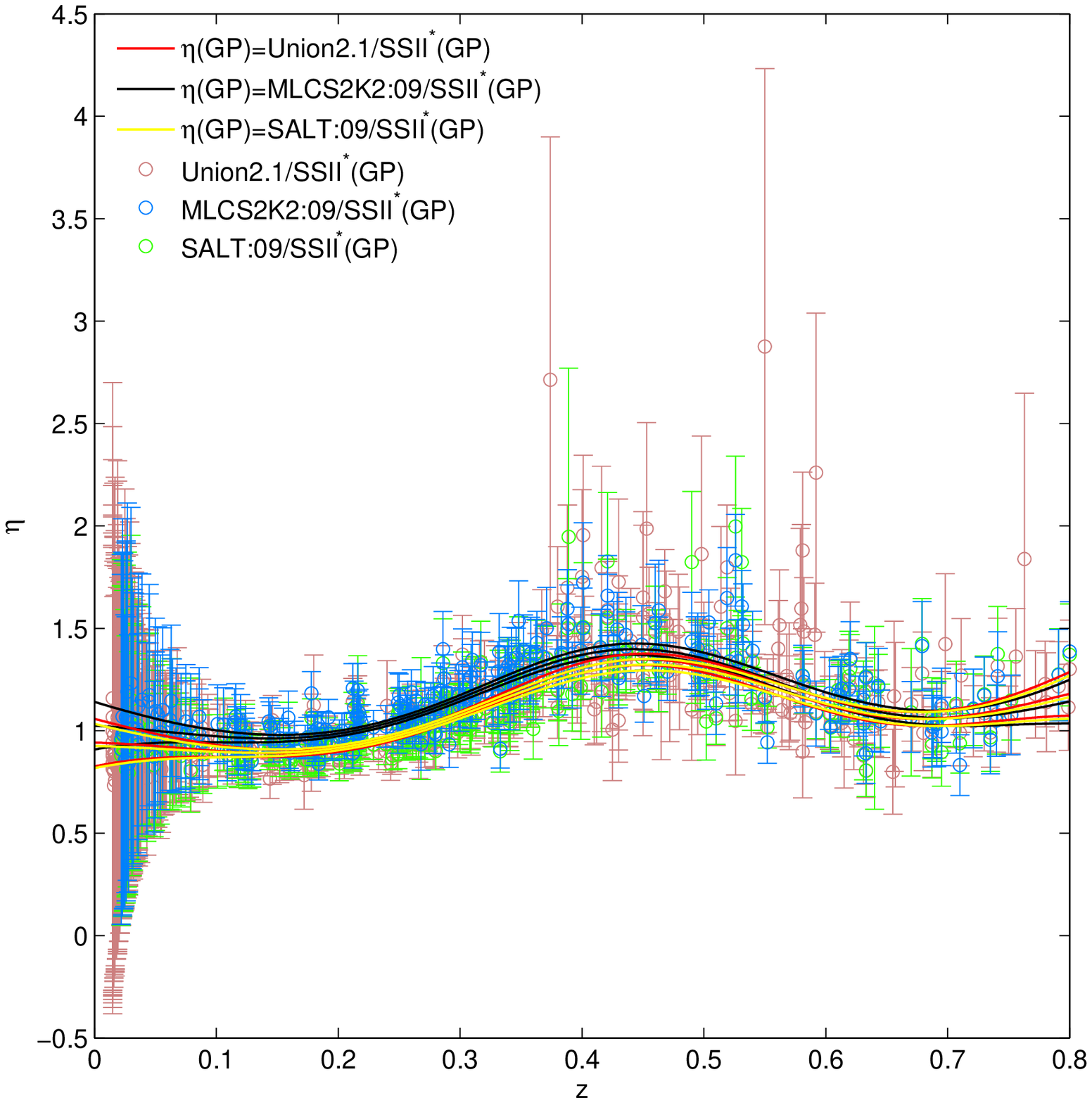}}\quad
     {\includegraphics[width=2.0in]{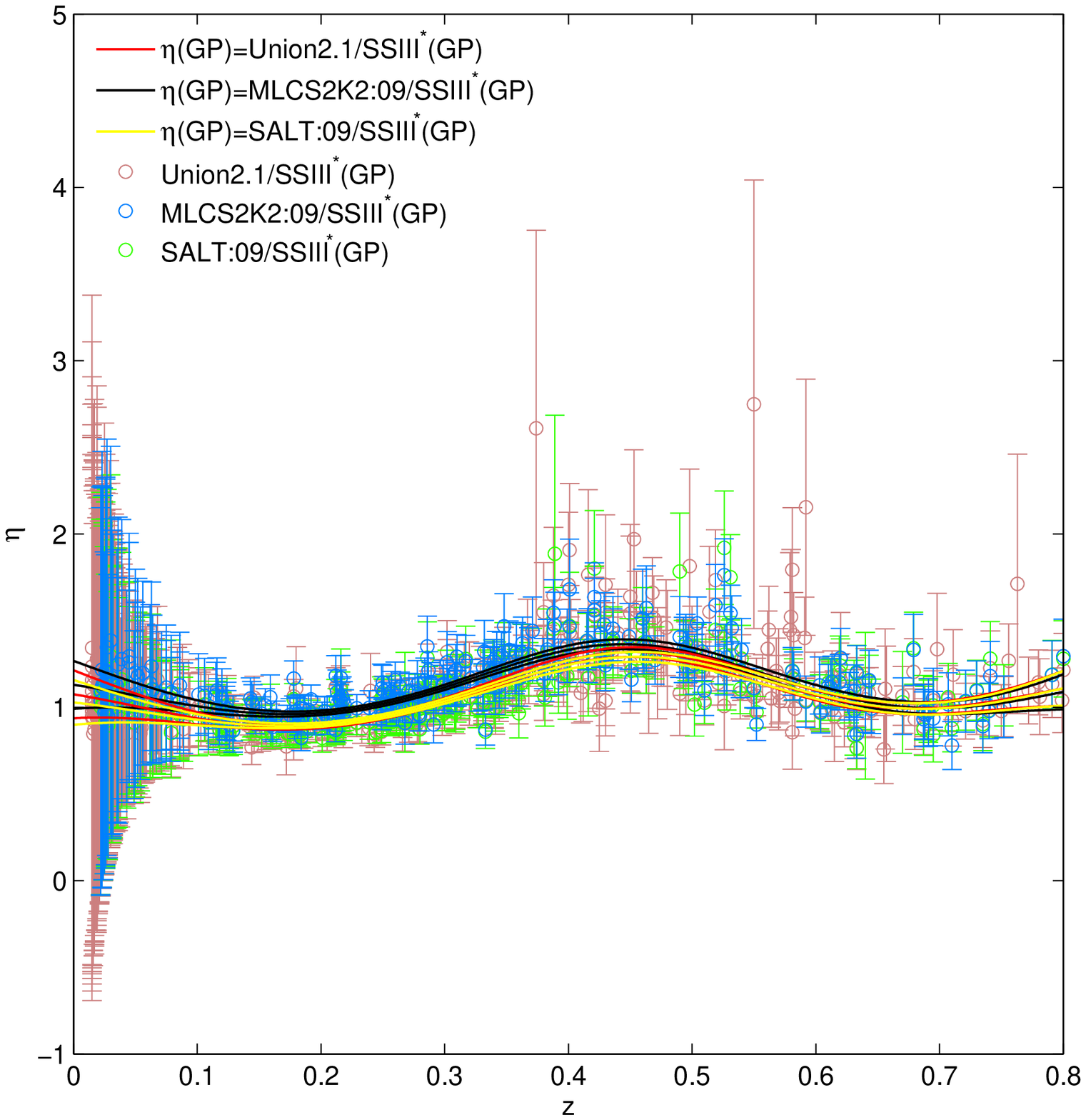}}\quad \\
  {\includegraphics[width=2.0in]{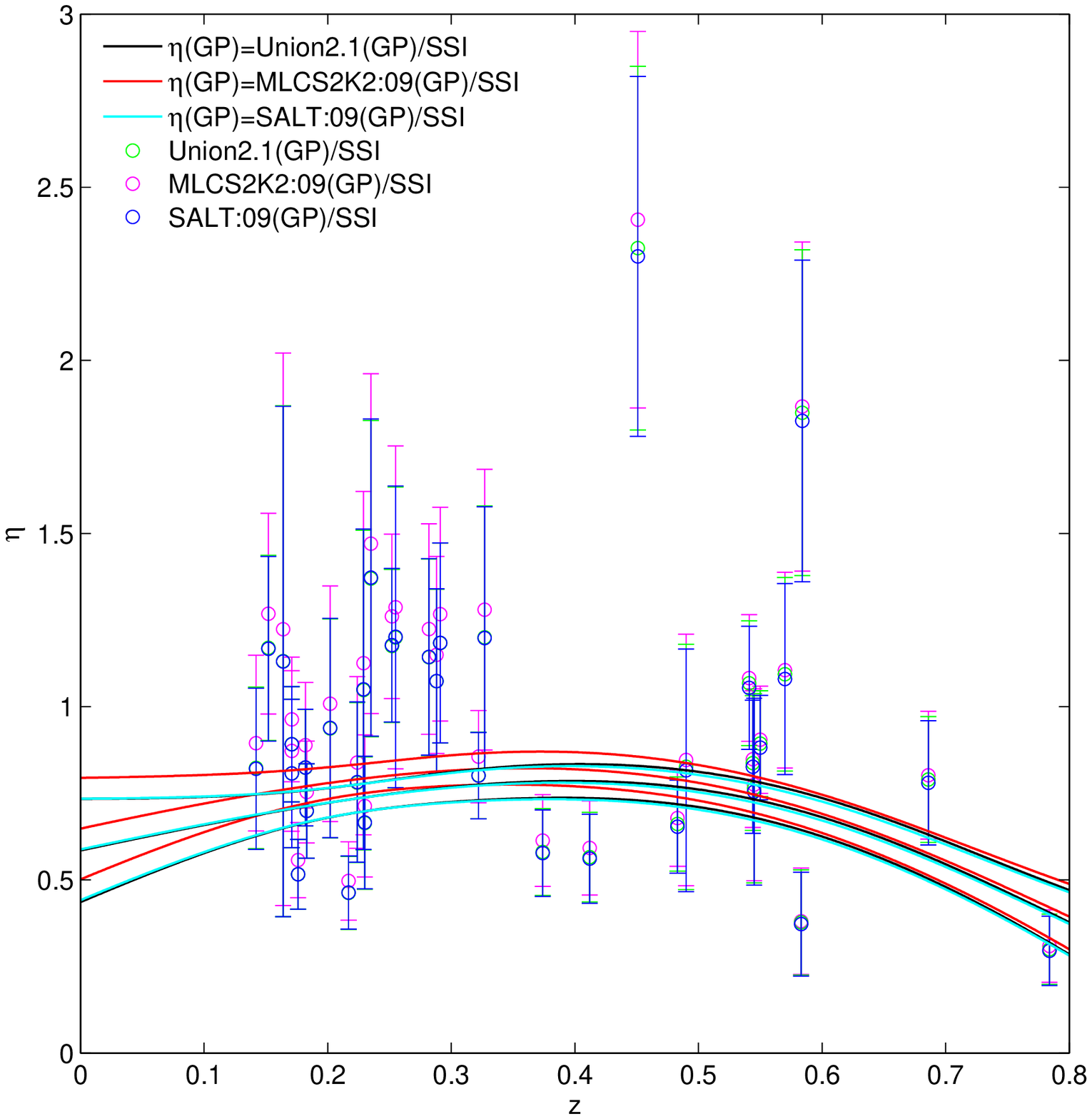}}\quad
    {\includegraphics[width=2.0in]{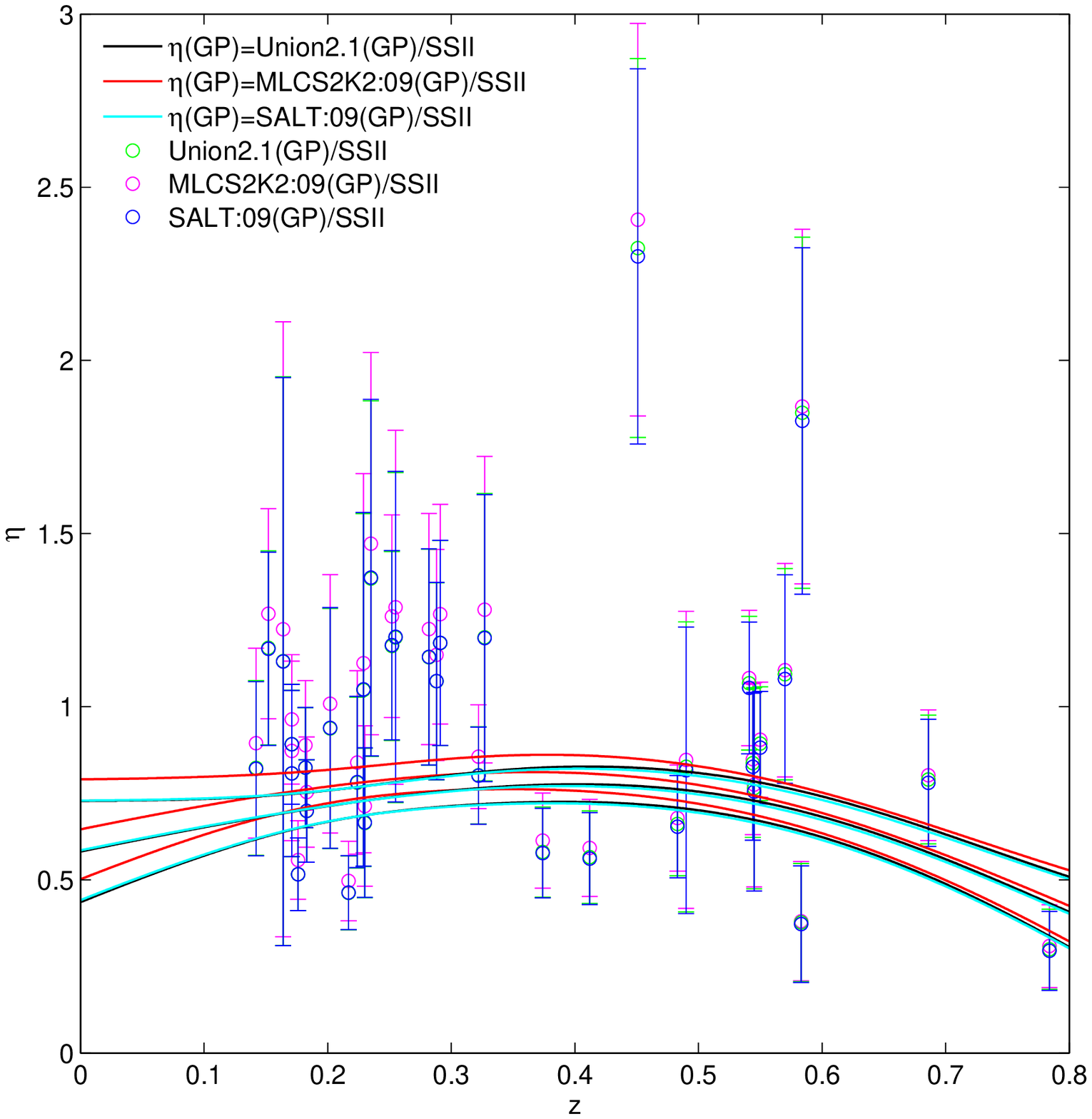}}\quad
     {\includegraphics[width=2.0in]{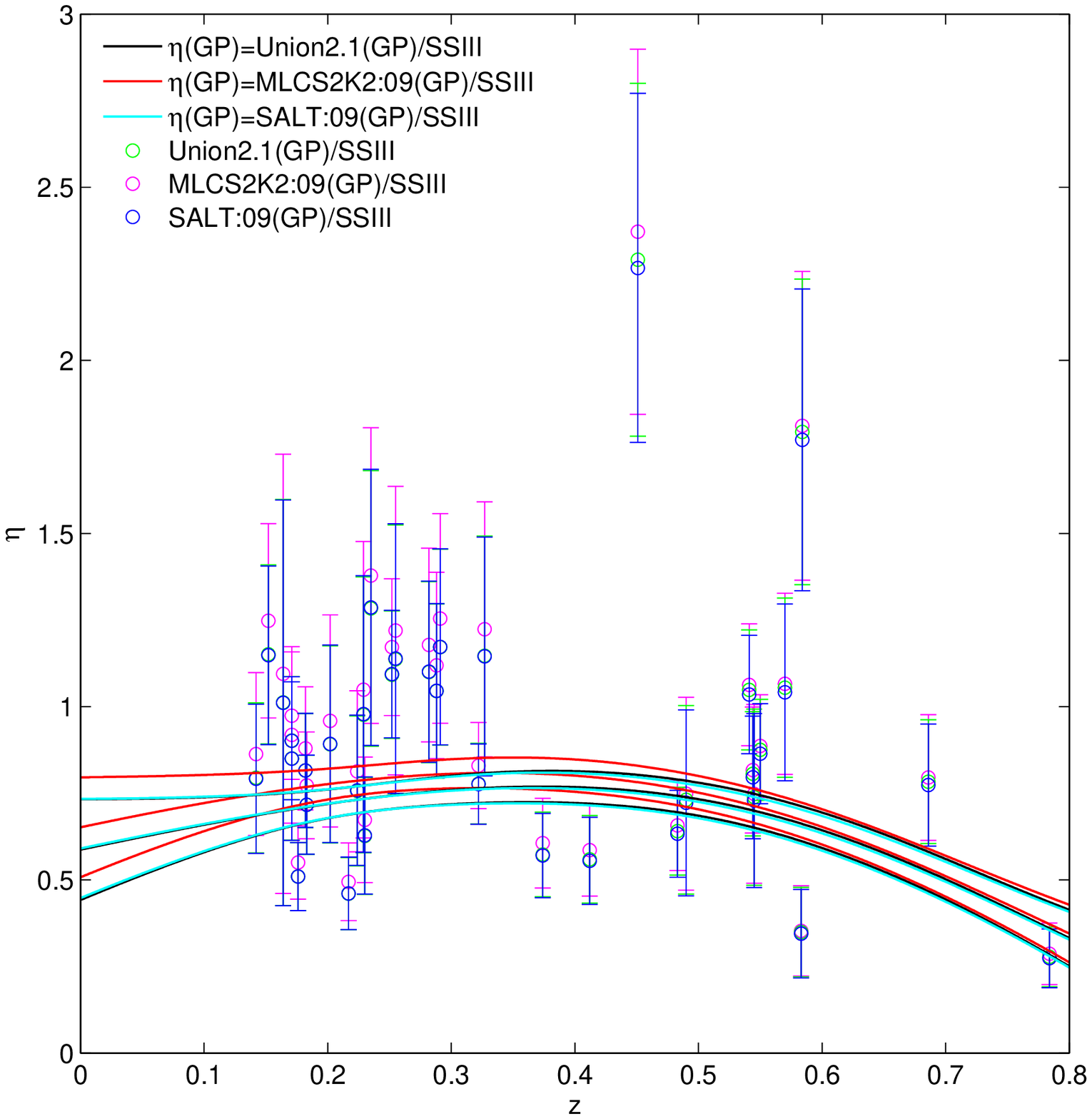}}\quad \\
       {\includegraphics[width=2.0in]{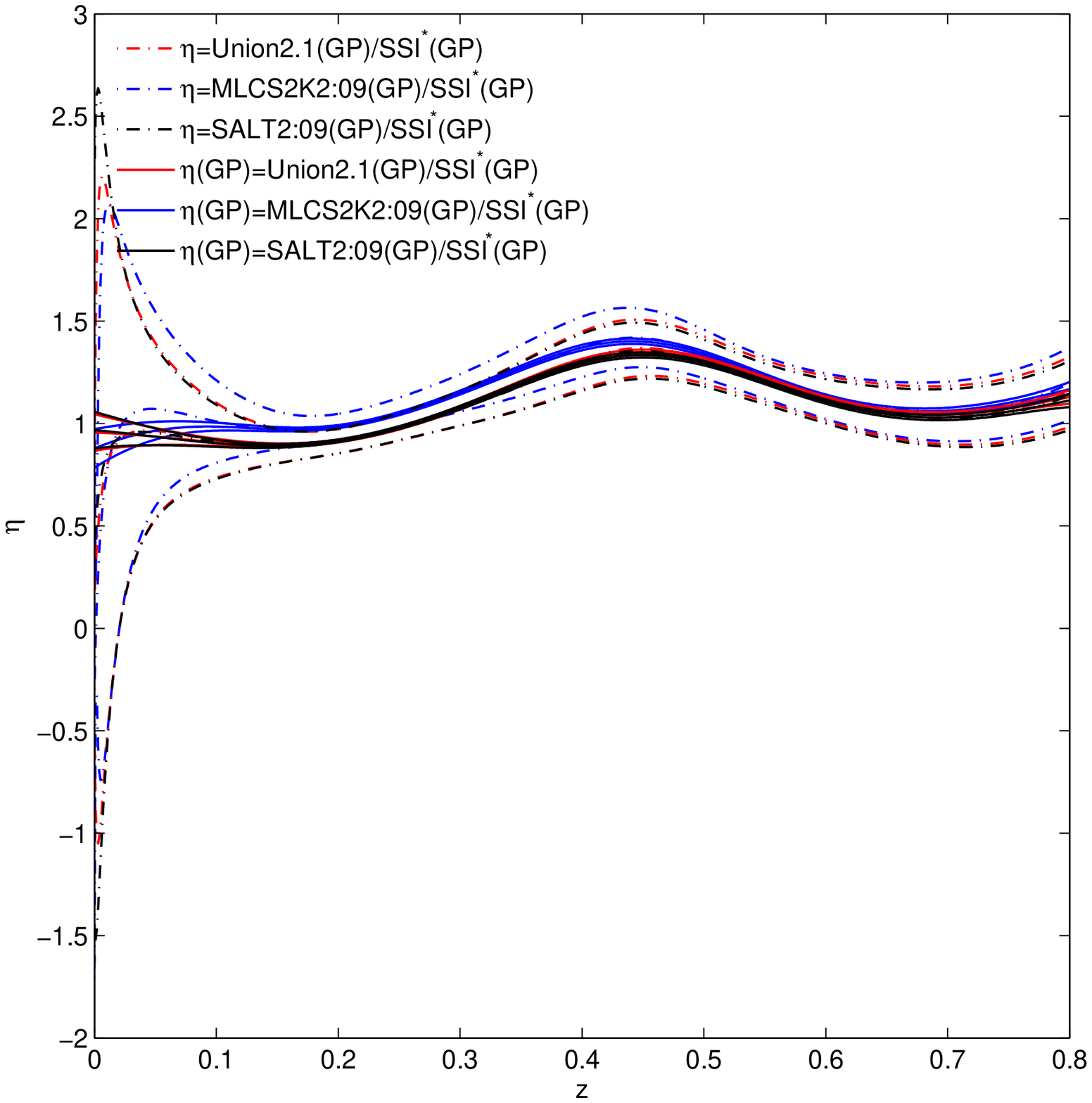}}\quad
    {\includegraphics[width=2.0in]{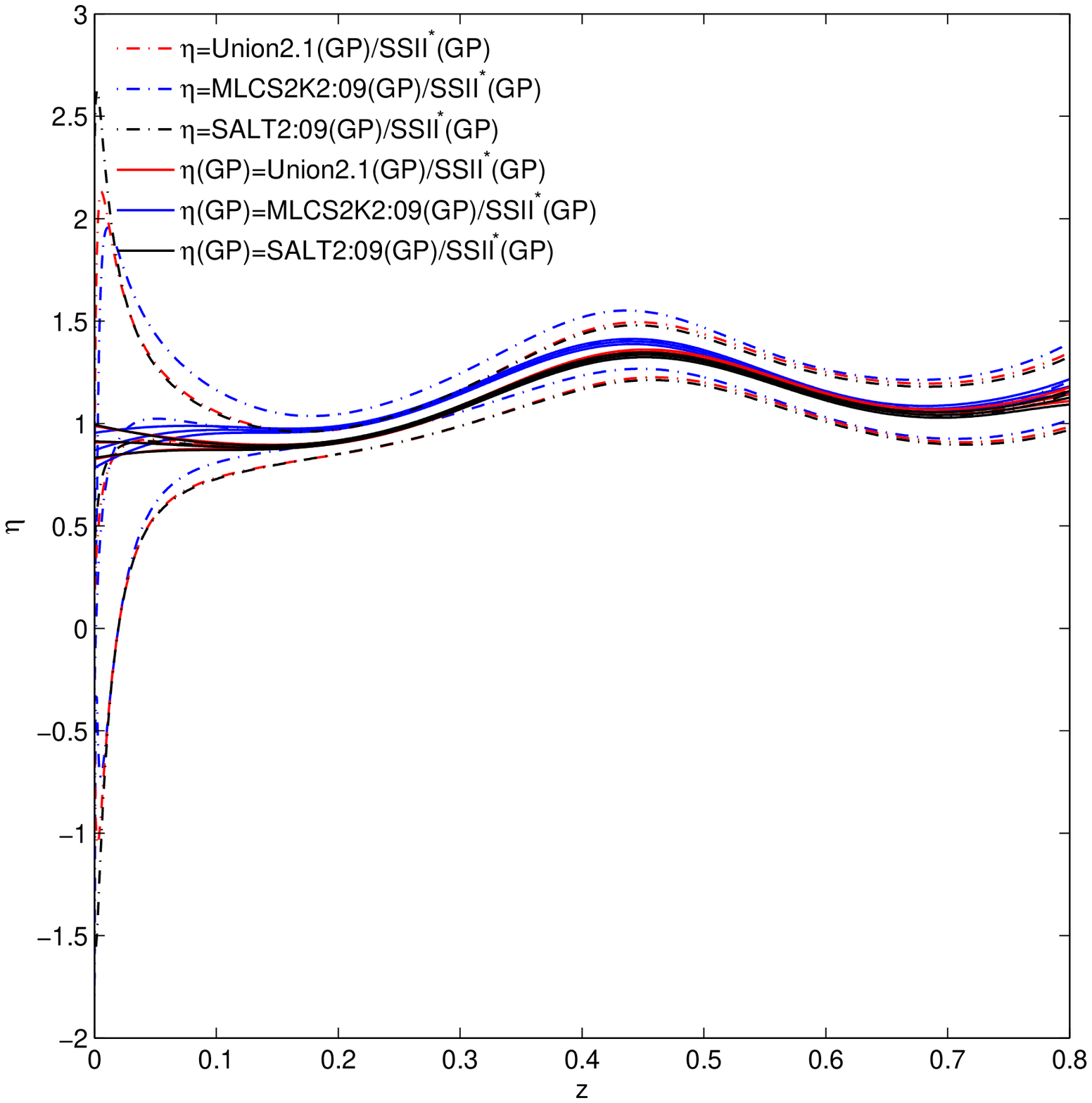}}\quad
     {\includegraphics[width=2.0in]{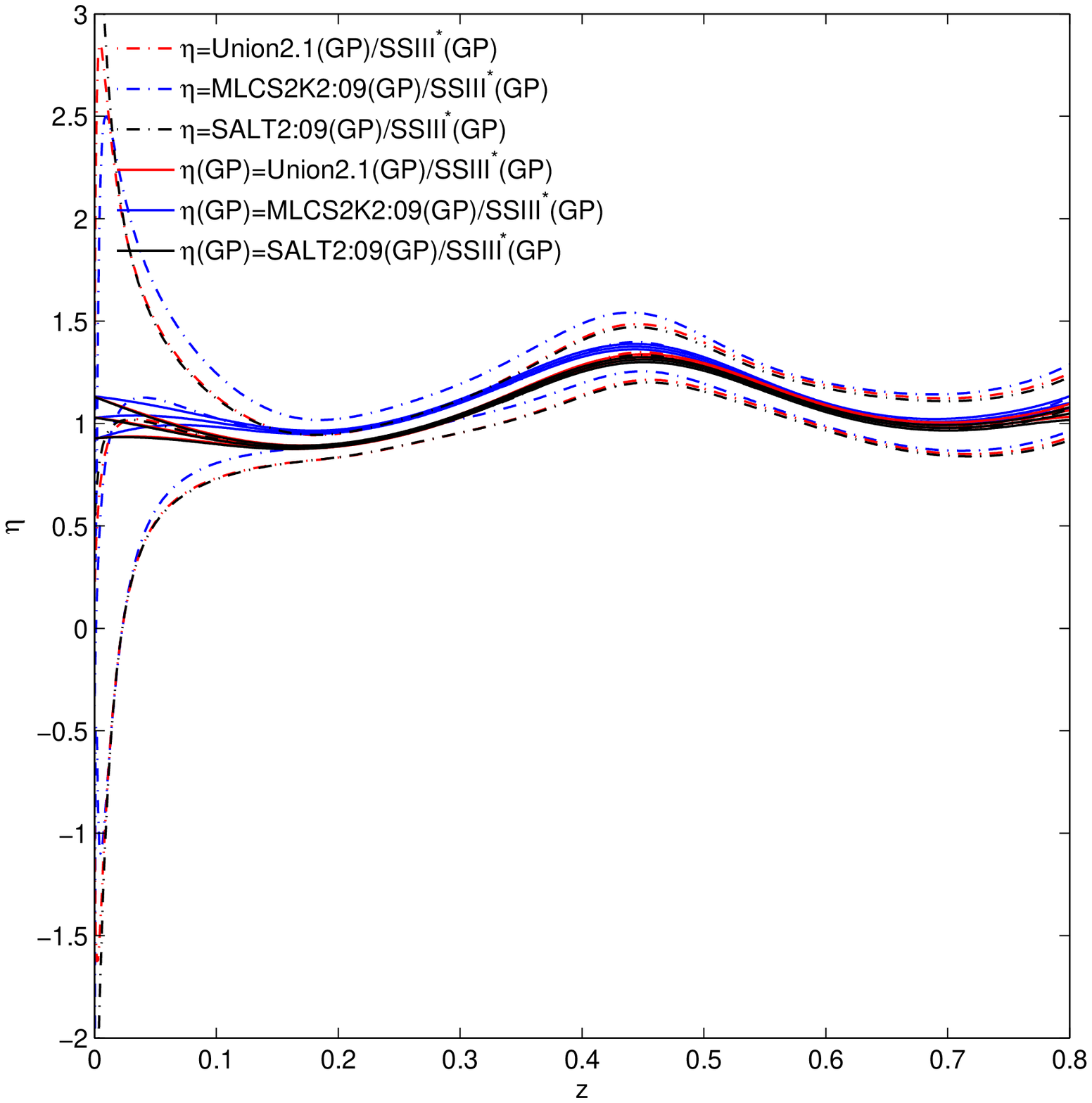}}\quad
    \caption{The upper, middle and lower  panels are  the GP results of $\eta$ of  LD(GP)/ADD, LD/ADD(GP) and LD(GP)/ADD(GP) respectively where the ADD data are derived from Ref.\cite{Bonamente:2005ct}. The points represent the observational data  with its error. The dashed lines are for $\eta= LD(GP)/ADD(GP)$. The solid lines are for the GP results.   }
  \label{2gpum}
\end{figure}

 As we have derived  the LD and  ADD at   same redshift by GP in Sec.\ref{sec4},   new
 $\eta$ samples could be constructed  from   the GP of LD and the observational data of ADD, or from  the GP of ADD and the observational data of LD, or 
from  both GP of ADD and LD  data. Then,
   the shape  of $\eta$  could be given out by using   GP again.   We present the new $\eta$ samples in Figure \ref{2gpall} and \ref{2gpum}. 
   
Generally speaking,    the GP results of $\eta=LD(GP)/ADD(GP)$ shown in Figure \ref{2gpall} and \ref{2gpum}  are not reliable.  Because they  are  far from DD relation at low redshift which could be tested by   Milky Way observation. This phenomenon may be  caused by the lack of observational data at low redshift.
We do not analyze the $\eta(GP)=LD(GP)/ADD(GP)$ results in the following. 
 And, most of the   MLCS2K2:09 related samples have  declining trends and  larger $\eta$ values than the Union2.1 and SALT2:09 related ones.

The left panels of Figure \ref{2gpall} present  the GP result of the  
 $\eta= $ LD(or GP)/$D_A|_{cosm}$(or GP) sample. As expected, the Union2.1/$D_A|_{cosm}$(GP)  and SALT2:09/$D_A|_{cosm}$(GP) (or Union2.1(GP)/$D_A|_{cosm}$ and SALT2:09(GP)/$D_A|_{cosm}$)  samples
 give out  flat $\eta$ trends and suggest   no violation of the DD relation.  
 The exact values of $\eta_{0}$ are shown in Table \ref{tab2}. These results are not a surprise because   the Union2.1, SALT2:09  and $D_A|_{cosm}$ data are based on cosmological parameter.  Meantimes,  It suggests the GP method is effective.

The middle panels in the vertical direction  of Figure \ref{2gpall} show  the GP results of the  
 $\eta= $ LD(or GP)/ES$|_{ell}$(or GP) sample.   Except the MLCS2K2:09/$ES|_{cosm}$(GP) data, the   derived  $\eta$ from GP which is related to  the ES$|_{ell}$ is  flat.
  The Union2.1(GP)/ES$|_{ell}$  and SALT2:09(GP)/ES$|_{ell}$  suggest that  no violation of the DD relation.  
 And,  the GP of  Union2.1/ES$|_{ell}$(GP)  and SALT2:09/ES$|_{ell}$(GP) data  obtain   a constant $\eta$ but $0.1$ below $\eta=1$. Especially, the GP of  MLCS2K2:09/$ES|_{cosm}$(GP) has a flat trend as well.  The exact value of $\eta_0$  could be  referred to Table \ref{tab2} as well.      It seems  the $\eta$ is flat and smaller than the DD relation   If ES$|_{ell}$ data  play main role.

 The right panels of Figure \ref{2gpall} show the GP results of the  
 $\eta= $ LD(or GP)/SS$|_{circ}$(GP) which are declining.  The   $\eta_0$ of     LD(or GP)/SS$|_{circ}$(GP) are listed in Table \ref{tab} as all which is  around $\eta=1$.   Basically,   the SS$|_{circ}$ should  have the similar result with the SS samples (SS\Rmnum{1}, SS\Rmnum{2}, SS\Rmnum{3})  from Ref.\cite{Bonamente:2005ct}. 
 But the SS samples have the bulge effect  around $\eta=1$ which are shown in Figure \ref{2gpum}.  The three different SS data with different error dealing method  do not show obvious difference.   And, the  bulge effect is hard to explained by   physics. 
 It is possible that  the bulge behaviors are caused by the GP of the SS sample which favors the data with smaller error bar.

\subsection{ Discussions}

 During all the GP reconstructions, no parameterization is assumed. The approach is fully model-independent, only the statistical quantities and covariance matrices are considered in the GP approach.

   Here, we concentrate on the test of  light-curve fitter. Nearly in all the comparison between MLCS2K2  and SALT2 fitters, the $\eta$ value from the MLCS2K2 fitter is larger than that from SALT2.
In Kessler's work,  the   constraining result  of Equation of State based on $w$CDM model   for MLCS2K2:09 is larger than  that based on SALT2:09 which indicates MLCS2K2:09 data itself have a  higher $D_{L}$.  
From this point of view, our results for DD relation are consistent with   Ref.\cite{Kessler:2009ys}.

 The connection between the validity of the DD relation and the morphology of galaxy cluster is investigated as well. In Refs.\cite{Cao:2011fw, Holanda:2011hh},  the verification of the DD relation seems to favor an elliptical shape, but for spherical model the validity of the DD relation seems only marginally compatible. 

Uzan \emph{et.al}  concluded there is no significant violation of the DD relation for a $\Lambda$CDM model by using ADD from X-ray surface brightness and Sunyaev-Zel'dovich effect measurements of galaxy cluster \cite{Uzan:2004my}. 
Our ES$|_{ell}$ sample includes the ADD data used by Ref.\cite{Uzan:2004my}.
    In  all the GP related to the ES$|_{ell}$ data,   the reconstructions of $\eta$ are nearly constant.  This phenomenon suggests the ES$|_{ell}$ favor the DD relation though the value are affected by LD data.  In another saying, if the  DD relation was correct, it would favor the galaxy morphology of elliptical one.

There is disagreement between our result and the result in Ref.\cite{DeBernardis:2006ii} where  De Bernardis \emph{et~al.}  gave out  no violating result of DD relation   and    found $\eta=0.97\pm0.03$ at $1\sigma$
 CL based on the SS. Though GP  
 has improved the error range of SS,  the SS\Rmnum{1}, SS\Rmnum{2} and SS\Rmnum{3}  related data display   bulge behaviors with  values smaller than $\eta=1$. 
The GP results of $\eta$ are not affected by the fake SS data  at $z=0$  because the $\delta z\leq 0.001$ sample and the LD(GP)/ADD sample also give out the bulge shapes.
  Our results based on the SS do not favor the DD relation.
When the spherical sample changed to SS$|_{circ}$, Figure \ref{0001} shows $\eta$  increases with redshift, while Figure \ref{2gpall}  show $\eta$  decreases with redshift.  The incompatible results are not reliable which may be caused by the large error. The spherical  samples do not favor the DD relation.


\section{The Summary}\label{sec6}
Based on Gaussian process,  we have introduced  new reconstructing approachs for DD relation. 
 The DD relation   relates the luminosity-distance with the angular-diameter-distance, and it has been widely used in cosmological models.  
 Checking the validity of DD relation with high accuracy  may provide probe of exotic physics.
GP is a non-parametric technique which could smooth the noise of data. In this letter, 
it has two effects in our $\eta$ reconstructions:  to reconstruct  the shape of $\eta$ and to get the LD and ADD data at the same redshift.

The related  data of  MLCS2K2 light-curve fitter   show  high $\eta$ values and  declining trends in most cases. In contrast, the related  data of  SALT2 light-curve fitters  (Union2.1 and SALT2:09)  are easier to get   constant   $\eta$. 
The GPs derived from  the Union2.1 and   $D_A|_{cosm}$ (the standard ruler) data show  $\eta$ is  close to $\eta=1$ which suggest GP is an effective method. 
The reconstructed  results related to $ES|_{ell}(GP)$ data show constant $\eta$   in most cases as well.
 And, for the SS related data, our three different methods of dealing asymmetric errors do not make obvious differences where all the related  GP results   have   bulge shapes. 
 And the varied behavior of the  SS$|_{circ}$ related  data   suggests its large error could not lead to  reliable results. Respect to galaxy cluster morphology,  the DD relation is favored by the elliptical one.

  \begin{table*}[t]
 \tiny
\begin{center}
\begin{tabular}{|llll|llll|}
\hline \multicolumn{4}{|c|}{The SS   } & \multicolumn{4}{c|}{The  ES  }\\
\hline Name & z& N(Union2.1/SS) & N(SALT:09/SS) &Name& z& N(Union2.1/ES) & N(SALT:09/ES)\\
 CL J1226.9+3332 &0.890  &- &- &MS 1137.5+6625  &0.784 & -&-\\
   MS 1054.5-0321  &0.826   &- &- & MS 0451.6-0305 &   0.550  &2 & 1\\
 RX J1716.4+6708  &0.813   & -&- &CL 0016+1609  &0.541    & -& -\\
 MS 1137.5+6625  &0.784  &- &- &RX J1347.5-1145  &0.451   &1 &1\\
   MACS J0744.8+3927  &0.686   & 2&2&Abell 370 &0.374    &1 &-  \\
  MACS J0647.7+7015 &  0.584   & 1&- & MS 1358.4+6245 & 0.327   &1 &1\\
  MS 2053.7-0449  &0.583   & 3&1 &Abell 1995 &0.322   &- &1\\
   MACS J2129.4-0741  &0.570   &3 &- &Abell 611 &0.288   &2 & 1 \\
   MS 0451.6-0305 &  0.550   & 2& 1 &Abell 697 &0.282   & -& -\\
   MACS J1423.8+2404  &0.545   &- &- &Abell 1835  &0.252   &2 &2 \\
   MACS J1149.5+2223  &0.544   &- &- & Abell 2261  &0.224   & -& -\\
 CL 0016+1609  &0.541    & 2&1 &Abell 773  &0.216    & 4& 3\\
  MACS J1311.0-0310  &0.490   &1 &1 &Abell 2163  &0.202   & 1&1 \\
  MACS J2214.9-1359 &0.483   &- &- &Abell 520  &0.202   & 1&1\\
 RX J1347.5-1145  &0.451   & 1&1 &Abell 1689  &0.183   & 1&-\\
      MACS J2228.5+2036  &0.412  & 1&- &Abell 665  &0.182   &1 &2\\
 Abell 370 &0.374    &1 &- &Abell 2218  &0.171    & 1& -\\
 MS 1358.4+6245 &  0.327   &1 &1 &Abell 1413 &0.142  & 2&- \\
 Abell 1995 &0.322   &- & 1 &Abell 2142  &0.091   &- &-\\
     ZW 3146 &  0.291   &2 &1 &Abell 478  &0.088   & 1&-  \\
Abell 611 &0.288   &2 & 1 &Abell 1651 & 0.084   & 1& -\\
 Abell 697 &0.282   &- &- &Abell 401 &0.074   & -&-\\
Abell 68 &0.255   & 1&- &Abell 399  &0.072   & -& 1\\
 Abell 1835  &0.252   & 2& 2 &Abell 2256  &0.058   &4 & -\\
 RX J2129.7+0005&   0.235  &- &- &Abell 1656 &0.023  & 7& 2 \\
   Abell 267  & 0.230  &- & -&-&-&-&-\\
 Abell 2111 & 0.229  &1 &1&-&-&-&- \\
  Abell 2261  &0.224   &- &-&-&-&-&- \\
  Abell 773  &0.217    &4 &3&-&-&-&-\\
  Abell 2163  &0.202   &1 & 1&-&-&-&-\\
  Abell 1689  &0.183   &1 & 1&-&-&-&-\\
 Abell 665  &0.182   &1 & 1&-&-&-&-\\
 Abell 2218  &0.176    & -&-&-&-&-&- \\
Abell 586  &0.171   & 2&-&-&-&-&- \\
 Abell 1914 &0.171   & -&-&-&-&-&- \\  
   Abell 2259 &0.164   &1 & 1&-&-&-&-\\
Abell 2204 &0.152  & 1& 1&-&-&-&-\\
     Abell 1413 &0.142  &2 &-&-&-&-&- \\
      Total Galaxy &- & 39& 22&Total Galaxy &- & 33& 17\\
          Removed Galaxy &- & 13& 20&Removed Galaxy &- & 8& 13\\
\hline
\end{tabular}
\end{center}
\caption[crit]{\small In the first column, the first and second rows are the name and redshift of the SS; the third and fourth  lines are the data number of  Union2.1/SS  and  SALT:09/SS samples based on the  $\delta z\leq 0.001$ criterion.  The informations in the second column are similar to the first one except that the ADD data is changed to the ES one. And, as  MLCS2K2:09/SS (MLCS2K2:09/ES) has the same number with SALT:09/SS (SALT:09/ES), we do not list the related information of MLCS2K2:09 for concise. }\label{tab}
\end{table*}
 

\section*{Acknowledgements}

I am grateful  for the  anonymous referee's important comments. 
This work was supported by    the National Natural Science
Foundation of China  under grant Nos.10935013 and 11175270.



\end{document}